\documentclass[twocolumn,tighten,twocolappendix]{aastex631}
\usepackage{amsmath}
\usepackage{color}
\usepackage{comment}
\usepackage[T1]{fontenc}

\def\approxgt{\ifmmode \rlap{$>$}{}_{{}_{{}_{\textstyle\sim}}} \else%
$\rlap{$>$}{}_{{}_{{}_{\textstyle\sim}}}$\fi}

\def\approxlt{\ifmmode \rlap{$<$}{}_{{}_{{}_{\textstyle\sim}}} \else%
$\rlap{$<$}{}_{{}_{{}_{\textstyle\sim}}}$\fi}

\begin{document}

\shorttitle{AT2018fyk a binary tidal disruption event}

\shortauthors{Wen et al.}

\title{
AT2018fyk: Candidate Tidal Disruption Event by a (Super)massive Black Hole Binary}

\author[0000-0002-0934-2686]{S.~Wen}
\email{wensx@bao.ac.cn}
\affiliation{National Astronomical Observatories, Chinese Academy of Sciences, 20A Datun Road, Beijing 100101, China}
\affiliation{Department of Astrophysics/IMAPP, Radboud University, P.O. Box 9010, 6500 GL, Nijmegen, The Netherlands}

\author[0000-0001-5679-0695]{P.G.~Jonker}
\affiliation{Department of Astrophysics/IMAPP, Radboud University, P.O. Box 9010, 6500 GL, Nijmegen, The Netherlands}

\author[0000-0001-7821-9369]{A.J.~Levan}
\affiliation{Department of Astrophysics/IMAPP, Radboud University, P.O. Box 9010, 6500 GL, Nijmegen, The Netherlands}

\author[0000-0002-4562-7179]{D.~Li}
\affiliation{National Astronomical Observatories, Chinese Academy of Sciences, 20A Datun Road, Beijing 100101, China}

\author[0000-0002-4337-9458]{N. C. Stone}
\affiliation{Racah Institute of Physics, The Hebrew University, Jerusalem, 91904, Israel}

\author[0000-0001-6047-8469]{A. I. Zabludoff}
\affiliation{University of Arizona and Steward Observatory, 933 N. Cherry Ave., Tucson, AZ  85721}

\author[0000-0002-0588-6555]{Z. Cao}
\affiliation{Department of Astrophysics/IMAPP, Radboud University, P.O. Box 9010, 6500 GL, Nijmegen, The Netherlands}
\affiliation{SRON, Netherlands Institute for Space Research, Niels Bohrweg 4,
2333 CA, Leiden, The Netherlands}

\author[0000-0002-4043-9400]{T. Wevers}
\affiliation{Space Telescope Science Institute, 3700 San Martin Drive, Baltimore, MD 21218, USA}
\affiliation{European Southern Observatory, Alonso de Córdova 3107, Vitacura, Santiago, Chile}

\author[0000-0003-1386-7861]{D.R. Pasham}
\affiliation{Kavli Institute for Astrophysics and Space Research, Massachusetts Institute of Technology, Cambridge, MA, USA}


\author[0000-0002-8671-1190]{C.~Lewin}
\affiliation{Kavli Institute for Astrophysics and Space Research, Massachusetts Institute of Technology, Cambridge, MA, USA}

\author{E.~Kara}
\affiliation{Kavli Institute for Astrophysics and Space Research, Massachusetts Institute of Technology, Cambridge, MA, USA}


\begin{abstract}
The tidal disruption event (TDE) AT2018fyk has unusual X-ray, UV, and optical 
light curves that decay over the first $\sim$600d, rebrighten, and decay again around 1200d.
We explain this behavior as a one-off TDE associated with a massive black hole (BH) \emph{binary}. The sharp drop-offs from $t^{-5/3}$ power laws at around 600d 
naturally arise when one BH interrupts the debris fallback onto the other BH. The BH mass $M_\bullet$ derived from fitting X-ray spectra with a slim disk accretion model and, independently, from fitting the early UV/optical light curves, is smaller by two orders of magnitude than predicted from the $M_\bullet$--$\sigma_*$ host galaxy relation, suggesting that the debris is accreted onto the \emph{secondary}, with fallback cut off by the primary. Furthermore, if the rebrightening were associated with the primary, it should occur around 5000d, not the observed 1200d. 
The secondary's mass and dimensionless spin is $M_{\bullet,{\rm s}}=2.7^{+0.5}_{-1.5} \times 10^5 M_\odot$ and $a_{\bullet,{\rm s}}>0.3$ (X-ray spectral fitting), while the primary's mass is $M_{\bullet,{\rm p}}=10^{7.7\pm0.4}M_\odot$ ($M_\bullet$-$\sigma_*$ relation).
An intermediate mass BH secondary is consistent with the observed UV/optical light curve decay, i.e., the secondary's outer accretion disk is too faint to produce a detectable emission floor.
The time of the first accretion cutoff constrains the binary separation to be $(6.7\pm 1.2) \times 10^{-3}~{\rm pc}$. X-ray spectral fitting and timing analysis indicate that the hard X-rays arise from a corona above the secondary's disk. The early UV/optical emission, suggesting a super-Eddington phase for the secondary, possibly originates from shocks arising from debris circularization.

\end{abstract}


\section{Introduction} \label{sec:intro}

When a star passes too close to a supermassive black hole (SMBH), the strong gravitational field can cause the star to be torn apart in what is known as a tidal disruption event \citep[TDE;][]{Hills1975,Rees1988}. A significant fraction of the resulting debris is then accreted onto the SMBH, causing a bright flare of radiation that can last for months to years. The mass fallback of the debris typically follows a power law, decaying at late times as  $t^{-5/3}$ \citep{Phinney1989}. The energy dissipated by the fallback debris powers a luminous flare that in principle enables the measurement of the SMBH mass and even spin
\citep{Mockler19,Ryu2020,Wen2020,Mummery+23}, although existing efforts at parameter estimation all involve significant assumptions. For this reason, TDEs have evolved into a powerful tool for investigating the properties of SMBHs.

When two galaxies merge, the two SMBHs will sink to the centre of the newly formed galaxy due to dynamical friction \citep{Begelman1980}. It is unclear if the BH binary shrinks all the way down to $\approx$0.1--0.01 pc where gravitational wave radiation determines the orbital evolution \citep{Milosavljevi2003}. Proposed processes that allow the binary SMBHs to evolve down to such small separations are: friction caused by the presence of a gaseous disk, a triaxial stellar distribution, and the presence of a third SMBH of similar size due to a subsequent merger \citep{Blandford1974,Merritt2004,Merritt2005,Mayer2007,Lodato2009,Vasiliev2015}. 
The existence of merging binary SMBHs emitting gravitational wave radiation is a major prediction underlying ESA-NASA's {\it LISA} mission \citep{LISA2023} and Chinese {\it TianQin/Taiji} mission \citep{Tianqin2021}.
The underlying population of low-mass SMBH binaries relevant for {\it LISA} remains unclear, although recent results from Pulsar Timing Arrays do lend support for merging SMBHs at higher masses \citep[]{Antoniadis2023,Agazie2023}.

When two SMBHs orbit each other at distances of a parsec or less, the TDE rate is enhanced by a combination of the Kozai-Lidov mechanism \citep{Ivanov2005} and chaotic three-body scatterings \citep{Chen2011}.  The TDE rate can be temporarily enhanced more than a hundred-fold, implying that multiple TDEs can be seen from the same galaxy on timescales of decades or less \citep{Chen2011,Wegg2011}. Some theories predict that overall about $10\%$ of the observed number of TDEs are associated with a supermassive black hole binary (SMBHB) \citep{Chen2011,Wegg2011}. This could explain why TDEs prefer host galaxies that experienced a burst of star formation that ended within the last $\sim$Gyr \citep{Arcavi2014,French2016,Graur2018}; such ``E+A'' or ``post-starburst'' galaxies (PSBs) often have disturbed morphologies indicative of galaxy-galaxy interactions and mergers.

Aside from indirect evidence related to statistics of host galaxy properties, the TDE light curves themselves can provide information on the question whether SMBHs merge. If one SMBH in a binary pair with a separation 
$\sim {\rm mpc}$ disrupts a star, the first stages of the TDE proceed as if the other SMBH were not present. However, due to the 
time-dependent gravitational field of the binary SMBH, some of the 
stellar debris is captured by the binary companion, truncating the mass fallback and thus the radiation 
emitted by accretion onto the SMBH responsible for the tidal disruption. The truncation time scale is a measure of the orbital period of the binary SMBH \citep{Liu_2009, Coughlin2019}.


A stellar disruption by the SMBHB \emph{secondary} is also possible and would be the exclusive channel for a TDE if the primary SMBH mass exceeds $10^8M_\odot$, which would swallow instead of disrupt stars \citep{Hills1975}. Although the probability of disruption by the secondary is low (proportional to the mass ratio) compared to that of the primary \citep{Coughlin2017}, the event rate associated with the secondary is shown to be significantly increased by the eccentric Kozai-Lidov (EKL) mechanism \citep{Kozai1962,Lidov1962,Naoz2016}, as reported by \cite{Mockler2023}. Such events may be more distinguishable observationally compared to cases of disruption by the primary. They not only exhibit enhanced variability in the light curve due to more pronounced perturbations from the primary, but may also produce huge discrepancies in SMBH mass estimates, specifically between estimates (i) based on emission from the TDE accretion disk, and (ii) those from host galaxy scaling relationships. 

The H$\alpha$ and H$\beta$ absorption indicates no ongoing star formation, and the lack of significant H$\delta$ absorption suggests that the host galaxy of AT2018fyk does not belong to the ``E+A'' galaxy class \citep{Wevers_2019}. Furthermore, the study of the emission lines shows that the host galaxy is not an AGN \citep{Wevers_2023}. AT2018fyk was initially classified as a TDE with $M_\bullet=10^{7.7\pm0.4} M_\odot$ from its host spectral properties \citep{Wevers_2019, Wevers_2021} and later reclassified as a repeating partial TDE (pTDE) due to the observed late-time rebrightening \citep{Wevers_2023}. In this paper, we propose an alternative explanation:
that AT2018fyk is consistent with stellar disruption by a SMBHB, with the secondary playing a crucial role.

Throughout this paper, we assume a flat concordance cosmological model ($\Lambda$CDM) with parameters $H_0=67.4~\rm {km ~s^{-1}~Mpc^{-1}}$ and $\Omega_m=0.315$ \citep{Planck2018}. The redshift of the host of AT2018fyk is $z=0.059\pm0.0005$, corresponding to a luminosity distance of $D_L=274$ Mpc. The paper is organized as follows.  We describe the (archival) optical/UV emission data and X-ray emission data in Section \ref{sec:data}. We show the results from analysing the UV/optical and X-ray emission in Section \ref{Results}.  In Section \ref{sec:binary}, we explain AT2018fyk as a SMBHB TDE. Finally, we summarize our conclusions in Section \ref{sec:Conclusions}.

\section{Data and Analysis} \label{sec:data}

\subsection{X-ray Spectra}

\begin{table*}
\centering
\caption{ A log of the {\it Swift} and {\it XMM-Newton} observations of AT2018fyk used in this paper. The number of source counts in the photon 0.3–10 keV energy band is given. The spectra are background subtracted. For {\it XMM} epochs, the exposure time and counts are shown in order of cameras pn, MOS1, and MOS2, respectively.
}
\begin{tabular}{cccccccc}
\hline
Spectrum & date  & Exposed time [ks] & Counts \\
\hline
{\it Swift} 1
           &2018/09/22-2018/10/17   &16.9  & 446 \\
{\it Swift} 2
           &2018/10/18-2018/11/21   &19.5  & 742 \\
{\it Swift} 3
           &2018/11/23-2019/01/08   &13.6  & 375 \\
{\it Swift} 4
           &2019/03/20-2019/05/20   &25.7  & 2419 \\
{\it Swift} 5
           &2019/05/23-2019/08/17   &20.7  & 1930 \\ 
{\it Swift} 6
           &2019/08/21-2019/10/23   &9.6  & 432 \\
{\it Swift} 7
           &2019/10/27-2020/01/09   &18.3  & 1267 \\
{\it Swift} 8
           &2022/03/25-2022/08/07   &44.2  & 556 \\
\hline
{\it XMM} 1
           &2018/12/09   &23.1/26.8/25.3  & 10372/2219/1791 \\
{\it XMM} 2
           &2019/10/27   &30.2/41.8/42.7  & 22325/6443/6416 \\
{\it XMM} 4
           &2022/05/20   &6.2/12.6/12.5  & 1213/528/637 \\
\hline
\end{tabular}
\label{Tab:swift}
\end{table*}

We reduce {\it Swift} X-ray data \citep{Evans2009} obtained for AT2018fyk over the period Sept.~22, 2018 to Aug.~7, 2022. We group and average the spectra into 8 epochs. In addition, we use 3 out of 4 epochs of the {\it XMM-Newton} (EPIC/PN) \citep{Jansen2001} X-ray data (see Table~\ref{Tab:swift} for a journal of the data used).

We use the online {\it Swift} XRT tool to  extract the time-averaged spectra\footnote{\href{https://www.swift.ac.uk/user_objects/}{https://www.swift.ac.uk/user-objects/}}. We centre on the source using the default {\tt Single pass} method. The {\tt Max attempts} and {\tt Search radius} are set at default values of 10 and 1, respectively. We do not select the {\tt Super-soft source} option and use the default {\tt all valid} Grade range. For the X-ray light curve, we choose the {\tt binning method} as ``observation'', allowing for upper limits and Bayesian bins. 

For the {\it XMM-Newton} data, we run the default {\sc SAS} v20 (20211130) tools under the HEASOFT {\sl ftools} software version 6.30 to extract source spectra and to filter the data. The EPIC pn detector \citep{Struder+01,Turner+01} provides the highest count rates but we also use the MOS1 and MOS2 data where available. We filter the pn data for periods of enhanced background radiation by deselecting episodes where the 10--12 keV detection rate of pattern 0 events is higher than 0.4 counts s$^{-1}$. We filter the MOS1 \& MOS2 data for periods of enhanced background radiation by deselecting episodes where the detection rate of pattern 0 events at energies $>10$~keV is higher than 0.35 counts s$^{-1}$. The effective exposure time for each observation after filtering is given in Table~\ref{Tab:swift}. The observations are done with the pn and MOS1\&2 detectors in Large Window ({\it XMM}~1) or Full Frame ({\it XMM}~2 \& 4) mode, respectively. We do not use the {\it XMM}~3 observation obtained on April 13, 2020 (MJD 58952.20), as the source was undetected \citep{Wevers_2021}.


\subsection{NICER Photometry}
{\it NICER} data was analyzed in the exact same manner as described in \cite{Wevers_2023} taking into account the possible contamination from the foreground Oxygen contamination. Additional details about the data reduction and screening thresholds can be found in \citealt{pashcow}.

\subsection{UV/Optical Photometry}

We reduce the UV/optical Telescope (UVOT; \citealt{Roming05}) data using the {\tt uvotsource} task, extracting fluxes using the standard 7 arcsec radius circular aperture. The emission in the UV bands has brightened by a factor of 10 compared to the host galaxy brightness, while the brightness of the B and V filters remains consistent with the inferred host galaxy brightness \citep{Wevers_2021}. We therefore do not include these latter two filters in our analysis. 

\subsection{Hubble Space Telescope Imaging}
The location of AT2018fyk was imaged by the Hubble Space Telescope on August 19, 2021 as part of programme 16239 (PI Foley). At this epoch, observations were obtained in F225W and F275W with exposure times of 780 and 710 seconds, respectively. Since only two dither positions were obtained in each filter, there remain numerous cosmic-ray hits which cannot be fully removed. However, a source at the location of AT2018fyk is clearly detected in both drizzled images obtained from the MAST archive (given the small number of dithers, a re-processing of the data to drizzle to a smaller pixel scale is not plausible). We measure the source magnitude in these epochs in a small (0.12 arcsecond) aperture centred on the position, obtaining magnitudes of F225W=24.48 $\pm$ 0.04 and F275W = 22.70 $\pm$ 0.02, when corrected for encircled energy. The use of a larger (0.32 arcsec) aperture gives magnitudes approximately 0.4 and 0.2 magnitudes brighter (again after correction for encircled energy), indicating there may be some extended underlying emission.

\section{Results} 
\label{Results}

\subsection{X-ray spectral fitting} \label{sec:X-ray}

Throughout this paper, we fit the spectra using {\sc XSPEC} version 12.12.0 \citep{Arnaud96}, applying Poisson statistics (\citealt{Cash79}; {\sc C-stat} in {\sc XSPEC}). We quote parameter errors at a $1\sigma$ $(68.3\%)$ confidence level (CL), using the method $\rm{Statistic} = \rm{Statistic_{best-fit}} + \Delta\,Cstat$ \citep{Arnaud96} and assuming $\Delta {\rm Cstat} = 1.0$ and $\Delta {\rm Cstat} = 2.3$ for single and two parameter models, respectively \citep{Wen_2021}. We use the slim disk model, {\tt slimd} \citep{Wen_2022}, to model the disk spectrum and the thermally comptonized continuum model, {\tt thcomp} \citep{Zdziarski_2020}, to model the hard part of the spectrum. The priors of their parameters are listed in Table~\ref{Tab:prior}. We set a prior on the inclination as $[2^\circ, 60^\circ]$, where $0^\circ$ represents a face-on orientation. When the disk is nearly edge-on ($>60^\circ$), a low accretion rate disk spectrum can be well fitted by a high accretion rate disk model, due to self-shielding by the finite aspect ratio of the disk \citep{Wen2020}. Edge-on TDEs typically exhibit additional absorption due to the sight line crossing outflowing material or intersecting the thick disk \citep{Dai2018,Wen2020}. However, there is no evidence for the presence of additional absorption in addition to the (Galactic) absorption parameterized by $N_{\rm H}=1.15\times10^{20} \mathrm{cm^{-2}}$ \citep{Wevers_2021}. Therefore, we enforce an upper limit on the disk inclination of $60^\circ$. We provide more details about the {\tt slimd} model and its capability to constrain disk parameters in Appendix \ref{app:2}.

\begin{table*}
\centering
\caption{Priors on the parameters of the {\tt slimd} and {\tt Thcomp} models
}
\begin{tabular}{cccccccccc}
\hline
 {\tt slimd } &$M_\bullet$ $[M_\odot]$  & $a_\bullet$ & $\theta$ [deg] & $\dot m$ [Edd] & {\tt Thcomp} & $\tau$ & $T_e$ [keV]\\
\hline
&[$10^3,9\times 10^7$]
           & [-0.998, 0.998]  &[2, 60] & [0.05, 100] & & [-100, -0.01] &[0.1, 300] \\
\hline
\end{tabular}
\label{Tab:prior}
\end{table*}

\subsubsection{{\it Swift} 1 Spectrum: Inner Accretion Disk}

\citet{Wevers_2021} showed that the source starts in a thermally dominated X-ray spectral state, which later develops into a power-law dominated state $\sim$100–200 days after the UV/optical peak. When analysing the time-averaged {\it Swift} spectra, we found that the power-law component becomes increasingly important after the date 2018/10/17. As a result, we designate the time-averaged spectrum from 2018/09/22 to 2018/10/17 as {\it Swift}~1. We fit the spectrum with a slim disk model ({\tt slimd}) \citep{Wen_2022}, a thin disk model ({\tt diskbb}), and a simplified but general disk model valid in the Wien emission limit ({\tt tdediscspec}) \citep{Mummery2021tde,Mummery+23}, each attenuated by an absorption model {\tt TBAbs} with a Galactic column density of N$_H$ = 1.15$\times 10^{20}$ cm$^{-2}$ (the same as in \citealt{Wevers_2021}). Table~\ref{Tab:s1} lists the fit results. See Figure~\ref{fig:spectra} for the spectrum with the best-fit slim disk model overlaid. For the slim disk model ({\tt slimd}), the free parameters are the BH mass ($M_\bullet$), the dimensionless BH spin ($a_\bullet$), the disk inclination ($\theta$), and the disk accretion rate $\dot m$ (in units of Eddington accretion rate, $\dot M_{\rm Edd}=1.37\times 10^{21} M_6~{\rm kg~s}^{-1}$, where $M_6=M_\bullet/(10^6M_\odot)$). We find that the spectrum can be well fitted with a $\rm {Cstat/d.o.f} \approx 1.0$ for all the three models. These results show that the early X-ray spectrum originated from the inner part of an accretion disk. The inferred values for the disk bolometric luminosity and $M_\bullet$, as determined through the slim disk model fits, are $1.4^{+4.0}_{-0.8}\times 10^{43}$ erg/s and $0.5^{+1.5}_{-0.3}\times10^5 M_\odot$ at $1\sigma$ CL, respectively. The disk bolometric luminosity is consistent with the Eddington luminosity of $L_{\rm Edd}\approx 2.5\times 10^{43}$ erg/s for a BH mass of $M_\bullet=2\times10^5M_\odot$, indicating that the disk is in a super-Eddington phase (also supported by the high accretion rate of {\it Swift}~1 in Table \ref{Tab:s1}). The best-fit unabsorbed X-ray luminosity is $L_{\rm X}=7.5\times10^{41}$ erg/s (for the band 0.3--10 keV), which implies that the ratio $L_{\rm X}/L_{\rm d}$ is in the range of $\approx 1-13\%$.

\begin{table*}
\centering
\caption{The best-fit parameter values and their $1\sigma$ errors from X-ray spectral fits to the {\it Swift}~1 time-averaged spectrum. We fit the spectrum with either the fit-function {\tt TBAbs $\times$ slimd },  {\tt TBAbs $\times$ diskbb}, or {\tt TBAbs $\times$ tdediscspec}. For the {\tt TBAbs} model, we fix the N$_H$ to the value  1.15$\times$ 10$^{20}$ cm$^{-2}$. Note that, the Eddington accretion rate is defined as $\dot M_{\rm Edd}=1.37\times 10^{21} M_6~{\rm kg~s}^{-1}$.
}
\begin{tabular}{cccccccc}
\hline
 {\tt TBAbs $\times$ slimd } &$M_\bullet$ $[10^5 M_\odot]$  & $a_\bullet$ & $\theta~ [deg]$ & $\dot m$ [Edd] & $Cstat/d.o.f$ \\
\hline
&$0.5^{+1.5}_{-0.3}$
           & $0.4^{+0.6}_{-0.8}$  &$54^{+6}_{-52} $ & $62^{+38}_{-61}$ & 87.86/90 \\
\hline
{\tt TBAbs $\times$ diskbb } &  & T [eV] &  & norm & $Cstat/d.o.f$ \\
\hline
&
           & $150\pm5$  & & 
           $200\pm40$ &89.01/92 \\
\hline
{\tt TBAbs $\times$ tdediscspec } &  & $R_{\rm peak}$ [$10^{12}$ cm] & $T_{\rm peak}$ [$10^5$ K] & $\Gamma$ & $Cstat/d.o.f$ \\
\hline
&
           & $0.09\pm 0.02$  &  $14\pm 3$ &
           $1.5\pm1.2$ &87.90/91 \\
\hline
\end{tabular}
\label{Tab:s1}
\end{table*}

\subsubsection{Evolution of Power-Law Component}
\label{hardXray}

We fit the {\it Swift} and {\it XMM-Newton} X-ray spectra simultaneously, with the fit-function ({\tt TBAbs $\times$ Thcomp(slimd)}) consisting of a slim disk model and a thermally comptonized continuum \citep{Zdziarski_2020}, absorbed by a Galactic column density of N$_H$ = 1.15$\times$ 10$^{20}$ cm$^{-2}$. For the slim disk component ({\tt slimd}), we force $M_\bullet$, $a_\bullet$ and $\theta$ to have time-independent values, while allowing for different accretion rates $\dot m$ at each epoch. We do allow the (time-independent) $M_\bullet$, $a_\bullet$ and $\theta$ parameters to float freely and hence the values can be different from the values determined from the fit to the {\it Swift}~1 and {\it XMM}~2 spectra separately. We will discuss this in more detail in Section \ref{sec:ma} and Appendix~\ref{app:1}.  

\begin{figure*}
    \centering
    \includegraphics[width=\textwidth]{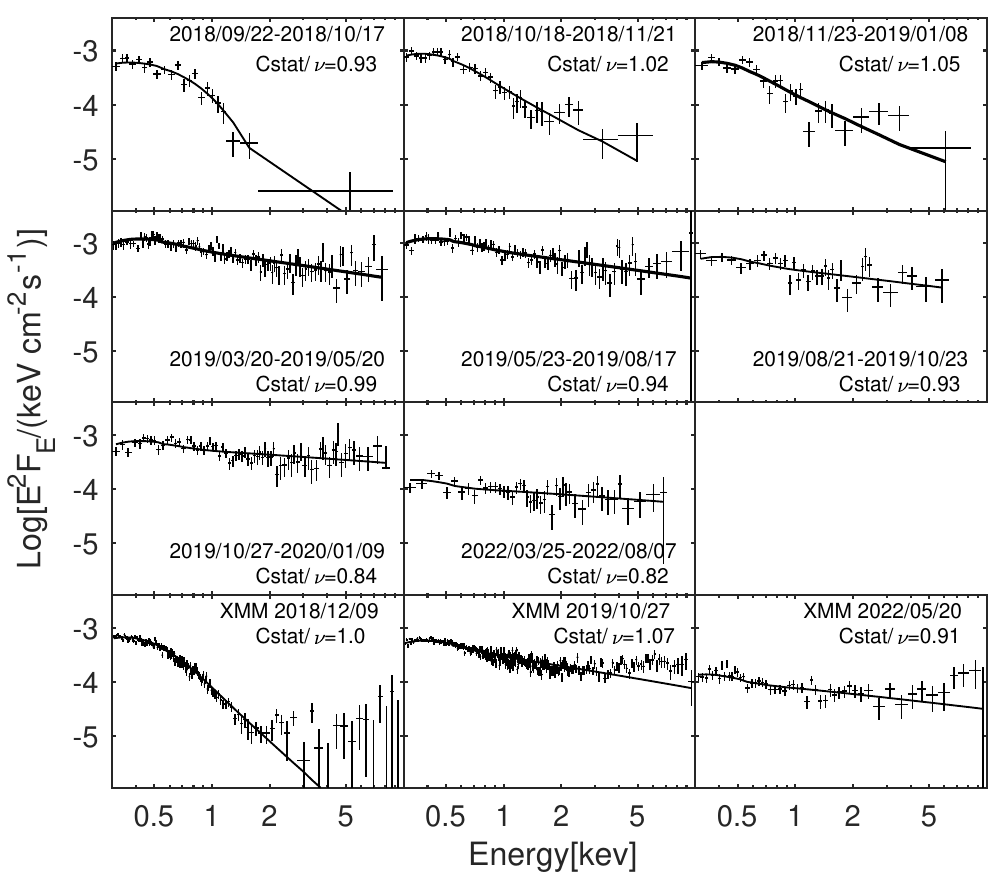}
    \caption{Simultaneous slim disk fits to eight {\it Swift} and three {\it XMM–Newton} 0.3--10 keV spectra for AT2018fyk from early ({\it Swift} 1) to late
({\it Swift} 8) times. The data are obtained over a four-year period. All the spectra are background subtracted.
For plotting purposes only, all the spectra are rebinned such that each bin has a $5\sigma$ significance. For the {\it XMM-Newton} spectra, we only plot the pn spectra. The fitted model and best-fit parameters are listed in Table~\ref{Tab:parameters}. }
\label{fig:spectra}
\end{figure*}

We first fit the spectra by allowing the different epochs to have a different optical depth and a different electron temperature for the {\tt Thcomp} model. We get a good fit with the total ${\rm Cstat/d.o.f.}=6749.3/6713$. We then fit the spectra by requiring all the epochs to have the same electron temperature. We also get a good fit with the total ${\rm Cstat/d.o.f.}=6755.7/6722$. The unfolded data and best-fit model spectrum for each epoch are shown in Figure~\ref{fig:spectra}. The best-fit parameters are listed in Table~\ref{Tab:parameters}. The $\Delta {\rm AIC}= -11.6$ \footnote{$\Delta{\rm AIC}=\Delta Cstat+2\Delta K$, where $K$ is the number of free parameters. Generally speaking, when comparing the best-fit statistics of two models, $\Delta {\rm AIC}=5$ and $10$ are considered to be strong and very strong evidence, respectively, against the weaker model.} \citep{Akaike1974}, obtained when comparing the fit where the electron temperatures are allowed to be different with the fit where they are tied together, shows that using the same corona electron temperature for all the epochs is strongly preferred by the data. We can see from Figure~\ref{fig:spectra} that the contribution of the hard component to the overall spectrum starts to become larger about one month after the discovery on 2018 September 8 (MJD = 58369.2), a trend which continues for the next three months. After epoch 4, the parameters of the best-fit spectral model stay roughly constant. This can, for instance, be seen from the fitted optical depth of each epoch. The optical depth grows from 0.1 at {\it Swift} 2 to about 0.3 at {\it Swift} 4. We interpret these results as the formation of a corona starting about one month after the peak of the UV emission. Its optical depth increases until it reaches a constant value in about 100 days. 

\cite{Wevers_2021} have also found that a power-law X-ray component emerges, with the source evolving from a soft state to a hard state, and they also attribute this power-law component to originate in a corona. In contrast to some X-ray binaries (XRBs) \citep{Remillard2006}, where the power-law strength increases simultaneously as the source luminosity decreases, AT2018fyk exhibits a disk luminosity that either increases or remains constant during the strengthening of the power-law component (refer to the lower panel of Figure \ref{fig:LRT}) \footnote{Note that if we assume the decreasing UV/optical emission associate directly with the source of the X-ray emission, then the source luminosity does decrease at the same time the power-law X-ray component become stronger.}. It is noteworthy that a similar power-law X-ray component manifests in the initial stages of certain TDEs, such as 3XMM J150052.0+015452 \citep{Cao2023}, AT2019azh \citep{Liu2022ApJ}, AT2019ehz \citep{van2021ApJ}, XMMSL1J0740 \citep{Saxton2017}, AT2020ocn \citep{Pasham2020}, and AT2021ehb \citep{Yao2022ApJ}. Some of the hard components observed in TDEs like 3XMM J150052.0+015452 and AT2020ocn have been explained by accounting for the effects of inverse comptonization by a corona, as proposed by \cite{Cao2023}. AT2021ehb exhibits a closely analogous evolution of its hard component, as illustrated in figure 15 of ~\cite{Yao2022ApJ}. The behavior of AT2021ehb's hard X-ray spectral component is attributed to the gradual formation of a magnetically dominated corona. The presence of a hard power-law X-ray component in the early spectra of TDEs may be associated with a super-Eddington accretion rate, diverging from the situation observed in certain Active Galactic Nuclei (AGNs) where the accretion rate is sub-Eddington \citep{Done2005}. If we assume that the power-law component comes from the corona, then the observed variations in coronal strength between different TDEs may reflect different magnitudes and topologies between the seed fields provided by the stellar debris, as magnetic initial conditions can strongly guide later evolution of accretion disc magnetic fields and coronal emission \citep{Liska+22}.

\begin{table*}
\centering
\caption{Best-fit values and the $1\sigma$ errors on the fit parameters. We fit the spectra simultaneously with a fit-function comprised of the model components ${\tt TBAbs \times Thcomp(slimd)}$. The Galactic column density, N$_H$, is fixed to a value of 1.15$\times 10^{20}$ cm$^{-2}$. The best-fit statistic is ${\rm Cstat/d.o.f.}= 6755.7/6722$.
}
\begin{tabular}{ccccccccc}
\hline
Spectrum & date & $10^5 M_\bullet$ $[M_\odot]$  & $a_\bullet$ & $\theta$ (deg) & $\dot m$ [Edd] & $\tau/(-1)$ & $T_e$ [keV] & ${\rm Cstat}/\nu$ \\
\hline
{\it Swift} 1 &2018/09/22-2018/10/17
           & $0.5^{+1.5}_{-0.3}$  &$0.4^{+0.6}_{-0.8}$  & $54^{+16}_{-52}$ & $62^{+38}_{-61.2}$ & -& -& 87.8/94\\
{\it Swift} 2 &2018/10/18-2018/11/21
          & $2.5^{+1.0}_{-1.0}$ & $0.9^{+0.1}_{-1.9}$  & $30_{-28}^{+30}$ & $1.0^{+0.7}_{-0.2}$ & $0.1^{+0.06}_{-0.06}$ & $300_{-10}$& 144.9/142\\
{\it Swift} 3 &2018/11/23-2019/01/08
           & $= M_{\bullet,2}$ & $= a_{\bullet,2}$  & $= \theta_{2}$ & $0.69^{+0.07}_{-0.05}$ & $0.11^{+0.07}_{-0.07}$ & $=T_{e,2}$& 124.3/118\\
{\it Swift} 4 &2019/03/20-2019/05/20
           & $= M_{\bullet,2}$ & $= a_{\bullet,2}$  & $= \theta_{2}$ & $1.5^{+5.1}_{-0.6}$ & $0.25^{+0.15}_{-0.15}$ & $=T_{e,2}$& 334.1/337\\
{\it Swift} 5 &2019/05/23-2019/08/17
         & $= M_{\bullet,2}$ & $= a_{\bullet,2}$  & $= \theta_{2}$ & $1.5^{+5.1}_{-0.6}$ & $0.25^{+0.15}_{-0.15}$ & $=T_{e,2}$& 294.6/313\\
{\it Swift} 6 &2019/08/21-2019/10/23
           & $= M_{\bullet,2}$ & $= a_{\bullet,2}$  & $= \theta_{2}$ & $0.67^{+0.14}_{-0.05}$ & $0.26^{+0.16}_{-0.16}$ & $=T_{e,2}$& 164.4/177\\
{\it Swift} 7 &2019/10/27-2020/01/09
           & $= M_{\bullet,2}$ & $= a_{\bullet,2}$  & $= \theta_{2}$ & $0.95^{+0.43}_{-0.34}$ & $0.31^{+0.19}_{-0.19}$ & $=T_{e,2}$& 254.3/302\\
{\it Swift} 8 &2022/03/25-2022/08/07
         & $= M_{\bullet,2}$ & $= a_{\bullet,2}$  & $= \theta_{2}$ & $0.23^{+0.05}_{-0.01}$ & $0.31^{+0.19}_{-0.19}$ & $=T_{e,2}$& 183.5/225\\
\hline
{\it XMM} 1 &2018/12/09
           & $= M_{\bullet,2}$ & $= a_{\bullet,2}$  & $= \theta_{2}$ & $0.73^{+0.07}_{-0.03}$ & $0.06^{+0.04}_{-0.04}$ & $=T_{e,2}$& 1084.9/1083\\
{\it XMM} 2 &2019/10/27
           & $4.5^{+1.0}_{-1.0}$ & $0.998_{-0.1}$  & $60_{-10}$ & $0.44^{+0.32}_{-0.03}$ & $0.23^{+0.14}_{-0.14}$ & $=T_{e,2}$& 3157.2/2950\\
{\it XMM} 4 &2022/05/20
           & $= M_{\bullet,2}$ & $= a_{\bullet,2}$  & $= \theta_{2}$ & $0.22^{+0.01}_{-0.01}$ & $0.27^{+0.17}_{-0.17}$ & $=T_{e,2}$& 925.7/1012\\
\hline
\end{tabular}
\label{Tab:parameters}
\end{table*}

\subsubsection{Fitted Black Hole Mass and Spin} 
\label{sec:ma}

We find that the fitted $M_\bullet$ differs moderately when fitting the different epochs (see Table~\ref{Tab:parameters}). We explore the variation in the best-fit values of $M_\bullet$ and $a_\bullet$ between different epochs in Appendix~\ref{app:1}. We found that the best-fit $M_\bullet$ of {\it XMM}~2 differs from the best-fit value found for the {\it Swift}, {\it XMM}~1 and {\it XMM}~4 spectra. {\it XMM}~2 has the highest number of counts in the spectrum. If we perform a joint fit that includes {\it XMM}~2, the fitted values of $M_\bullet$ and $a_\bullet$ are driven by the data of {\it XMM}~2. However, a strong power-law component is present in the {\it XMM}~2 spectrum, making the {\tt slimd} parameters less robust.  Therefore, we exclude the {\it XMM}~2 data from the joint fit that has as a goal to constrain the $M_\bullet$ and $a_\bullet$.

\begin{figure}
    \centering
    \includegraphics[width=0.5\textwidth]{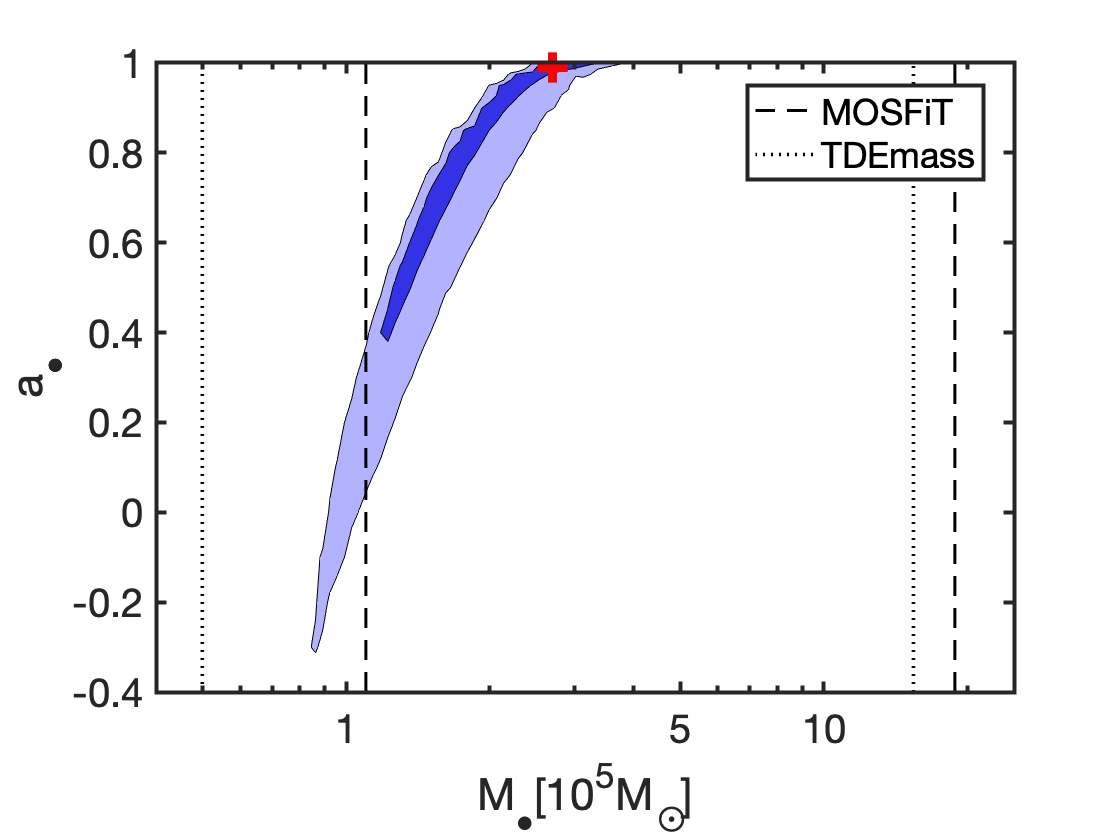}
    \caption{Constraints on $M_\bullet$ and $a_\bullet$ from simultaneously fitting a fit-function ${\tt TBAbs \times Thcomp(slimd)}$ to the X-ray spectra of all the {\it Swift} epochs, and the {\it XMM}~1 and {\it XMM}~4 spectra. The red cross shows the best-fit value of the parameters, while the dark and light blue regions denote the 1 and $2\sigma$ contours, respectively. We also plot the $1\sigma$ mass range inferred from modeling the UV/optical emission by the {\sc MOSFiT} and {\sc TDEmass} packages.}
    \label{fig:maep4}
\end{figure}

In order to get the strongest constraints on $M_\bullet$ and $a_\bullet$ possible with this data, we fit the spectra {\it Swift}~1-{\it Swift}~8, {\it XMM}~1 and {\it XMM}~4 simultaneously with the fit-function ${\tt TBAbs \times Thcomp(slimd)}$. Since we showed that the {\it Swift}~1 spectrum is consistent with just a disk spectrum, we disable the model component {\tt Thcomp} from the fit-function for {\it Swift}~1 by fixing the covering fraction factor to 0. For all the other epochs, we assume the same $M_\bullet$, $a_\bullet$ and $\theta$, but allow for a different $\dot m$ value. Except for {\it Swift}~1, we use the model {\tt Thcomp} to describe the hard component and assume the same electron temperature, but allow for a different optical depth at each epoch. Figure~\ref{fig:maep4} shows the constraints on $M_\bullet$ and $a_\bullet$. We obtain the best-fit value of $M_\bullet=(2.7^{+0.5}_{-1.5}) \times 10^5 M_\odot$ and $a_\bullet>0.3$ at a $1\sigma$ CL. As shown in the plot, the $M_\bullet$ inferred from the X-ray spectra fitting also agrees with the $M_\bullet$ measured by {\sc TDEmass} \citep{Ryu2020} and {\sc MOSFiT} \citep{Mockler19}, obtained through modeling the UV/optical emission of this event (see Section~\ref{UVMass} for details on UV/optical emission modeling).

We also estimate $M_\bullet$ with the {\it Swift}-1 fits using a thin disk and a model for a spreading TDE disk \citep{Mummery+23}. Assuming the disk terminates at the innermost stable circular orbit (ISCO) radius $r_{\rm ISCO}$, the thin disk model predicts the evolution of the maximum disk temperature as $T_{\rm eff, max} \approx 78 (\frac{r_{\rm ISCO}}{r_g})^{-3/4}(\frac{\dot m}{M_7\eta_{-1}})^{1/4}~\rm{eV}$ \citep{Shakura1973}, where $r_g=GM_\bullet/c^2$, $M_7=M_\bullet/10^7M_\odot$, $\eta_{-1}=\eta/0.1$, $G$ is Newton's constant, $c$ is the speed of light, and $\eta$ is the radiative efficiency). We get the mean effective temperature $T_{\rm eff,mean} \approx 2\sqrt{\frac{r_{\rm ISCO}}{r_{\rm out}}} T_{\rm eff,max}$ (for the case of $r_{\rm out} \gg r_{\rm ISCO}$) by integrating the flux across radii. Due to electron scattering in a hot atmosphere with finite temperature gradients, the real X-ray flux will be higher than the corresponding black-body flux \citep{Shimura1995}, resulting in a higher effective temperature (by a factor of $f_c$, the spectral hardening factor). 

For the case of $a_\bullet=1.0$ (where $r_{\rm ISCO}/r_g=1.0$ and $\eta=0.42$) and $\dot m \ge 1$ \footnote{For highly super-Eddington accretion rates, the bolometric luminosities increase over the Eddington limit by factors $\sim \ln(\dot m)$ \citep{Abramowicz2005}. The X-ray luminosities increase slower (even constant), due to a decreasing $f_c$ \citep{Wen2020}. Overall, we assume a constant $T_{\rm eff}$ for $\dot m>1$.} , by adopting $r_{\rm ISCO}/r_{\rm out}=1/20$ and $f_c=2.0$ (for the super-Eddington cases; \citealt{Davis2019}), we get $T_{\rm eff}\approx 50 M_7^{-1/4}~\rm{eV}$.  We note that this estimate is in good agreement with observations from ASASSN-14li, which exhibited a temperature of $T\sim 55$eV during the first 200 days post-discovery \citep{Miller2015}, with best-fit $M_\bullet=10^7 M_\odot$ and $a_\bullet=0.998$ (inferred from slim disk modeling; \citealt{Wen2020}), as well as CXO~J1348, where a constant black-body temperature of $\sim 90$ eV was detected early in the outburst, with $M_\bullet=10^{5.5\pm0.5}M_\odot$ (inferred from the $M_\bullet$--$M_{\rm bulge}$ relation; \citealt{Donato2014}). Given the $T_{\rm eff}\approx 50 M_7^{-1/4}~\rm{eV}$, the 150 eV inferred from {\tt diskbb} fitting for epoch 1 (see Table~\ref{Tab:s1}) indicates that $M_\bullet$ is on the order of $\sim10^5 M_\odot$. In addition, the TDE disk model-fits using {\tt tdediscspec} constrains $M_\bullet$ to be $(4.4\pm1.0) \times 10^5 M_\odot$, with a systematic error of $3.5\times 10^5 M_\odot$.  This systematic error arises from the uncertainty in the relationship between $R_{\rm Peak}$ and $M_\bullet$. The disk bolometric luminosity is found from {\tt tdediscspec} to be $(2.2_{-1.7}^{+5.3})\times 10^{43}$ ergs/s. The fitted $M_\bullet$ and the disk bolometric luminosity, obtained by applying Equations 13 and 19 of \cite{Mummery+23}, are consistent with the outcomes of the slim disk modeling.

In summary, multiple types of X-ray spectral fitting analysis reveal that the thermal X-ray emission originates from a disk surrounding a BH with a mass of approximately $2\times10^5 M_\odot$. Our fiducial multi-epoch X-ray spectral modeling using the slim disk model finds that the BH mass and spin are $(2.7^{+0.5}_{-1.5}) \times 10^5 M_\odot$ and $a_\bullet>0.3$ at $1\sigma$ CL, respectively. The corona that becomes unambiguously detected after the first epochs contributes to the X-ray spectra in the form of (hard) power-law emission. Its contribution grows, eventually stabilizing after about 100 days. The X-ray emission from this event shares similarities with that of IMBH TDEs, such as 3XMM~J215022.4$-$055108 \citep{Lin2018,Wen_2021} and 3XMM~J150052.0+015452 \citep{Lin2017,Cao2023}, which also exhibit high effective temperatures and prolonged super-Eddington decay phases. Notably, in the case of 3XMM~J150052.0+015452, its early hard X-ray component can also be explained by incorporating a corona. In the following section, we will present additional evidence supporting the association between the power-law X-ray emission and a corona for AT2018fyk. 

\subsection{UV/optical emission analysis}

\subsubsection{{\tt Swift} UVOT and HST observations}
\label{uvothst}

\begin{figure*}
    \centering
\includegraphics[width=1.0\textwidth]{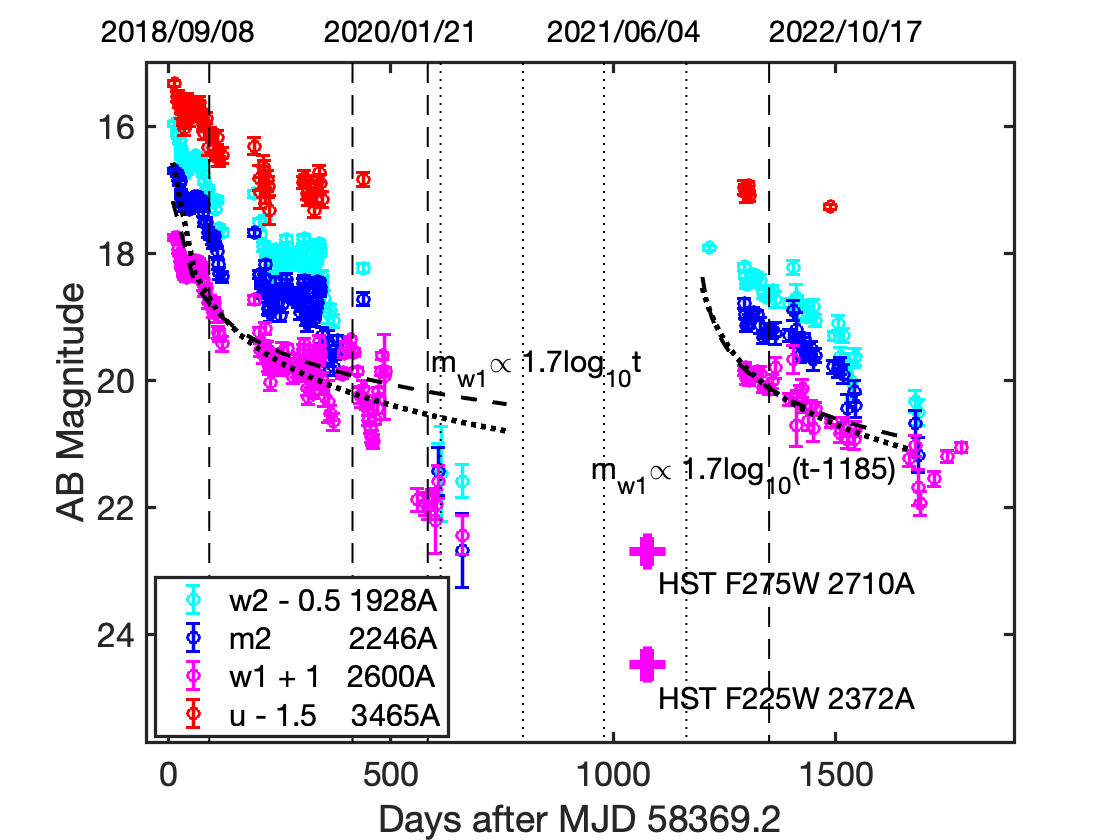}
    \caption{The {\it Swift/UVOT} and {\it HST} observations of AT2018fyk. The colored circles represent the { \it Swift/UVOT} observations, while the two pink pluses denote the {\it HST} observations. The dashed and dotted solid vertical lines denote the times of the observations of {\it XMM} and {\it eROSITA}, respectively. Over the time range between $t=583$ and 1164~d, the {\it XMM}~3 and the four {\it eROSITA} observations yielded non-detections of the source X-ray flux \citep{Wevers_2023}.  The black dashed lines depict the best power-law fit to the w1 light curve, while the black dotted lines show the best fit of $m_{\rm w1}\propto (9/4)\log_{10} t$ to the w1 light curve. This figure shows that the UV luminosity decays as the mass fallback rate for the first 500 days, decreases quickly and reaches the host galaxy level by $\sim561$ days, at least until the {\it HST} observation at $t$=1076~d. This is followed by another power-law decay after the rebrightening that started by $t\approxlt1216$ ~d.
    }
    \label{fig:UV}
\end{figure*}

Figure \ref{fig:UV} shows the evolution of the UV emission as a function of time. The magnitudes are not corrected for any extinction from our own Galaxy or the host galaxy. The UV observations at $\sim 600$ days and the HST observation on day $1076$ are consistent with the host galaxy's brightness, indicating that the TDE flare became significantly dimmer during these periods. The w2 and m2 magnitudes decline more rapidly compared to w1 around 600 days, suggesting a substantial reduction in the temperature of the source (see also the top panel of Figure \ref{fig:LRT}). The {\it HST} observation also supports a lower temperature on day $1076$: the magnitude of short-wavelength band F225W is significantly fainter than that of the long-wavelength band F275W, while m2 tends to be brighter than w1 in the period $\approxlt500$ days and after $\sim1200$ days. Additionally, X-ray observations by the {\it XMM-Newton} in particular observations {\it XMM~3}, {\it Swift}, and {\it eROSITA} satellite between the days 561 and 1163 show that the brightness of the source fell below the detection limit \citep{Wevers_2021,Wevers_2023}. This further supports the conclusion that the TDE flare experienced a significant decrease in brightness during those periods. 

To determine the decay rate of the UV light curve, we use a power-law function for the flux, ${\rm Flux} \propto t^n$, to fit the w1 observations during the initial 500 days. We obtain a best-fit index of $n=-1.70$, in good agreement with the classical mass fallback rate index of $-5/3$ \citep{Phinney1989}.  Even though the UV/optical luminosity fluctuates, it generally seems to follow the mass fallback rate. We also fit the rebrightening apparent in the w1 light curve with a power-law function for the flux, $ {\rm Flux} \propto (t-t_0)^n$, where we fix the index at the best-fit value from the first light curve fitting, $n=-1.70$. Here, $t_0$ represents the fitted peak date of the rebrightening. We obtain a best-fit value of $t_0=1185$~d, with $\chi^2/\nu= 54.3/30$. Notably, this value of $t_0$ is in line with the last X-ray non-detection from $\it eROSITA$ at  $t=1163.7$~d, and the first detection after the rebrightening in the UV was at $t=1216$~d \citep{Wevers_2023}. For comparison, we also show the theoretical partial disruption fallback decay rate, which is given by ${\rm Flux}\propto t^{-9/4}$ \citep{Coughlin2019}. The best-fit $\chi^2$ values for the early $\sim500$ days and the rebrightening $\approxgt 1200$ days are 3530.9 and 57.68, respectively. The first one is notably larger than the corresponding value of 1962.7, while the second one close to that of 54.3 obtained from the $n=-1.70$ fit.

Past studies show that TDE UV/optical light curves (after peak) decay initially as a power-law (typically $<350$ days), then reach a plateau. The initial power-law decay is often consistent with theoretical fallback rates, although the precise early-time emission mechanism remains unclear \citep{Loeb+97, Piran+15, Roth+20}.  Late-time emission in the plateau phase is thought to originate from the outer edge of a circularized accretion disk \citep{Ulmer99, vanVelzen+19,Hammerstein2022,Mummery2023b}. However, in AT2018fyk, both the decay during the first 500 days and the decay after the rebrightening are well-fit by power-laws, with no evidence of a plateau. The outer disk UV emission is thus probably fainter than the observed values. Using the slim disk model \citep{Wen22b}, we can calculate the expected $w1$ magnitude from the outer disk using the parameters derived from fitting the X-ray spectra. For a disk around a BH with $M_\bullet=2\times 10^5 M_\odot$, and an accretion rate of $\dot m=2$, we get $m_{w1}\sim 24$, which is indeed much fainter than the observed value of $16<m_{w1}<20$ and also far below the host brightness ($m_{w1}\sim 21$) \citep{Wevers_2019}. As the brightness of the disk in $w1$ is fainter than that of the host, we predict that the observed $w1$ emission will continue to decay following a power-law until it reaches the host level, which is expected to occur approximately 5000 days after discovery. However, if the star was disrupted by the secondary of a SMBHB, accretion to the primary BH would eventually interrupt the power-law decay. 

In general, both power-law decay phases are consistent with the idea that the UV/optical luminosity is dominated by the same emission process. Observations from {\it Swift}, {\it HST}, {\it XMM}, and {\it eROSITA} also support the fact that the TDE flare subsides at the very least, or even shuts off, between 561 and 1163.7 days. We attribute the rapid decrease in both UV and X-ray emission to a cutoff in the gas fallback stream and subsequent cessation of accretion onto the BH.

\subsubsection{Black-body fit to the UV/optical emission}

\begin{figure}
    \centering    \includegraphics[width=0.5\textwidth]{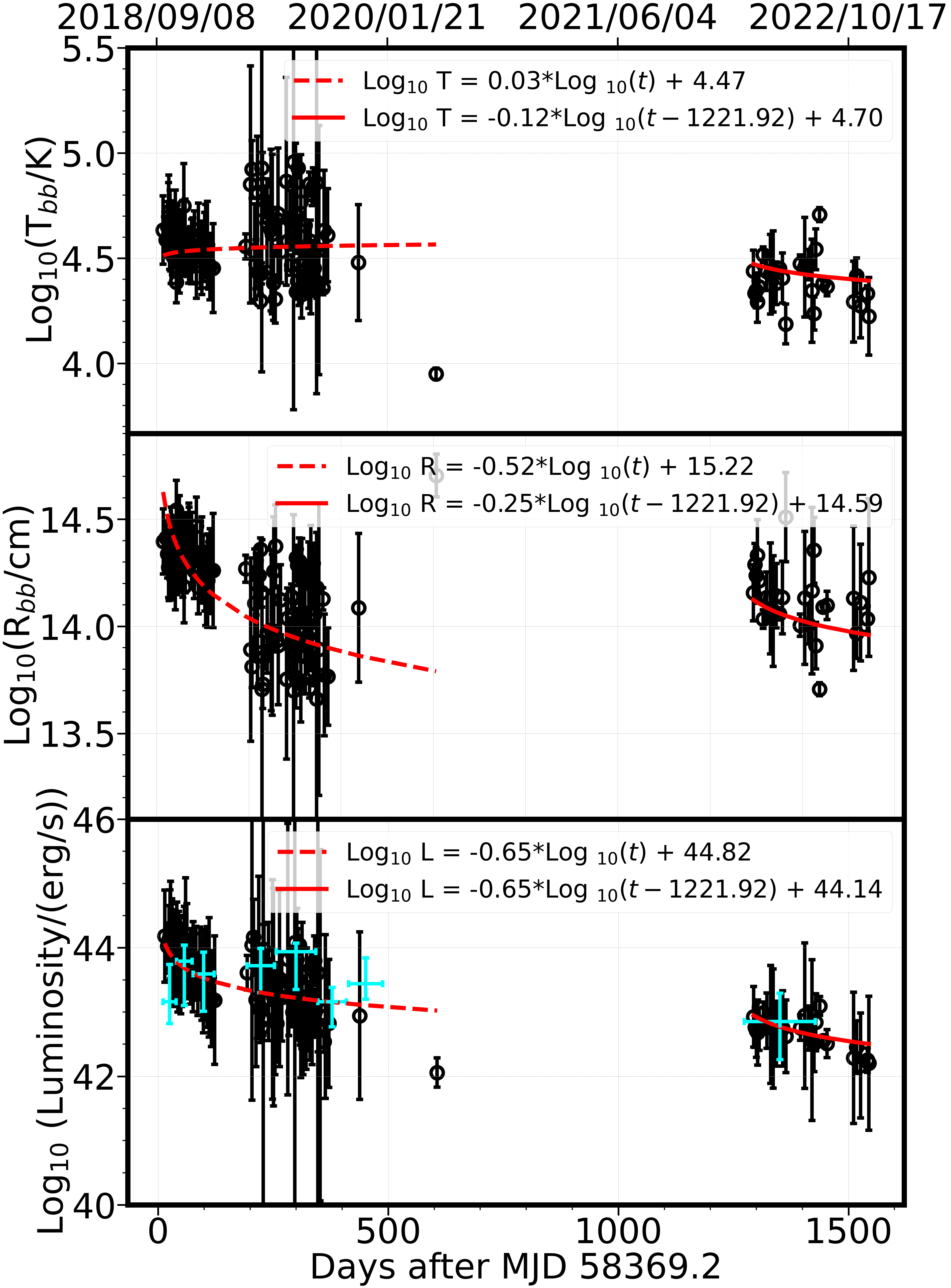}    
    \caption{Evolution of the black-body radius (top panel), temperature
(middle panel), and luminosity (bottom panel) over time. Black open circles denote the early and rebrightening epochs, respectively. The red dashed lines denote the best-fit power-law model for the early epoch, while the red solid lines represent the model for the rebrightening epoch. In the bottom panel, the cyan pluses show the disk bolometric luminosity as a function of time derived from X-ray spectral fitting of the 8 {\it Swift} averaged spectra.}
    \label{fig:LRT}
\end{figure}

The SED of the early power-law decay UV/optical emission in TDEs can often be well described by thermal black-body emission \citep{vanVelzen+11}. We therefore measure the black-body temperature and radius, as well as the luminosity, by fitting a single-temperature black-body model to the UV/optical data as a function of time. For all the UV/optical observations available, we subtract the host galaxy contribution as determined from SED fitting in \citet{Wevers_2021}.
The host galaxy subtracted fluxes are also corrected for Galactic extinction by employing the IDL code {\sc fm\_unred.pro} with $E(B-V) = 0.01$ \citep{Schlegel1998}, assuming the Milky Way (MW) extinction curve \citep{Gordon2003}. To carry out the black-body model fitting of the UV/optical data, only observations with positive host-subtracted fluxes in more than 2 filters are used\footnote{The host contribution dominates the emission in some observations, which might result in negative host-subtracted flux densities, especially in optical bands.}.  We use the flux density at a specific wavelength, $f_{\lambda} = 4\pi R_{\rm bb}^{2}D_L^{-2} B_{\lambda}(T_{\rm bb}, \lambda)$, to do the fitting. Here, $R_{\rm bb}$ and $T_{\rm bb}$ are the black-body radius and temperature, respectively, and $B_{\lambda}(T_{\rm bb}, \lambda)$ is the Planck law as a function of wavelength. We then calculate the black-body luminosity with the Stefan-Boltzmann law, $L_{\rm bb} = 4\pi \sigma_{sb} R_{\rm bb}^2  T_{\rm bb}^4$, where $\sigma_{sb}$ is the Stefan-Boltzmann constant.

Figure~\ref{fig:LRT} shows the black-body temperature, radius, and luminosity as a function of time. The relatively large uncertainties are due to the fact that the peak of the black-body spectra is not sampled by the observations. The evolution of the temperature and radius is consistent with the results of \citet{Wevers_2021} for both the early and the rebrightening epochs, e.g., constant temperature and slowly decreasing radius. 

We fit the luminosity with the equation $\log_{10} L_{\rm bb} = -\alpha \log_{10} (t-t_{0}) + C$, where $t_{0}$ is the peak day and $C$ is the luminosity at $t-t_{0} = 1$. Here, we set the zero point of $t$ at MJD 58369.2 days. Giving $t_0 = 0$~d for the early epochs, the best-fit parameters are $\alpha = 0.65 \pm 0.09$ and $C = 44.82 \pm 0.20$. During the rebrightening epochs, assuming $\alpha$ is the same as the early epochs, the best-fit $t_0$ and $C$ are $1221.92 \pm 41.19$~d and $44.14 \pm 0.08$, respectively. The luminosity decays as a power-law for both the early and rebrightening epochs, but with a smaller index of 0.69 when compared with the power-law decay index of $w1$. Integrating the best-fit over time, we obtain the total black-body energy released for the early and rebrightening epochs; they are $1.1\times 10^{51}$ ergs (close to the estimation of $1.4\times 10^{51}$ ergs from \citealt{Wevers_2019}) and $1.1\times 10^{50}$ ergs, respectively. 

If we adopted a radiation efficiency of $\eta=0.01$ (see our analysis in the next subsection), the radiated energy amounts to $0.06 M_\odot$, and $0.006 M_\odot$ for the early and rebrightening epochs, respectively, indicating $M_\star>0.13M_\odot$ assuming 50\% of the star's material is unbound and 50\% is accreted. This is in agreement within 1~$\sigma$ with the star's mass $M_\star=0.07^{+0.12}_{-0.04}M_\odot$ estimated from fitting the UV light curves  (see Table~\ref{tb:uv}). For the early epochs, the peak black-body luminosity and temperature are $10^{44.28\pm0.75}$ erg/s (in agreement with the estimate of $(3.0\pm0.5)\times10^{44}$ erg/s from \citealt{Wevers_2019}) and $10^{4.72\pm0.17}$ K, respectively.  Given the Eddington luminosity $L_{\rm Edd}\approx3.3\times10^{43}$ erg/s (for the case of $M_\bullet=2.7\times 10^5M_\odot$), the UV black-body luminosity is about 10 times the Eddington luminosity. It takes about 200 days for the UV/optical emission to decline to the Eddington luminosity. We also plot disk bolometric luminosities derived from the slim disk modeling of the {\it Swift} X-ray spectra. We fit each of the {\it Swift} spectra with the model ${\tt TBAbs \times Thcomp \times slimd}$ separately, and then calculate the unabsorbed disk bolometric luminosity with the disk parameters from the {\tt slimd} component. As shown in Figure~\ref{fig:LRT}, the BB luminosity of the UV/optical emission is brighter than the slim disk bolometric luminosity for the first epoch. 

It is possible to generate super-Eddington luminosities in the early UV emission. Shock-powered emission \citep{Piran+15} is not necessarily Eddington-limited, and neither is reprocessing in an unbound disk wind \citep{Metzger2016} if the central engine is itself super-Eddington.  The cooling envelope model \citep{Metzger2022} predicts that the bolometric luminosity is about $L=L_{\rm Edd}+ L_{\rm fb}$, where $L_{\rm Edd}$ is the disk Eddington luminosity and $L_{\rm fb}$ is produced by the fall back debris. We present observational evidence supporting the presence of a super-Eddington luminosity. Firstly, \cite{Wevers_2019} detected low-ionization Fe II emission lines about a month after discovery. These emission lines require an obscuring medium with significant particle density and optical depth, along with an energy source to heat the gas. This obscuring medium can be attributed to the super-Eddington outflow. The presence of these emission lines is consistent with our assumed prior on the slim disk inclination angle to our line of sight, $\theta<60^{\circ}$, since a sight line with $\theta>60^\circ$ implying we would be looking through the outflow, might have resulted in absorption lines. In addition, the X-ray flux exhibited a gradual increase, eventually reaching a plateau around 200 days \citep{Wevers_2023}. The gradual increase in the X-ray flux (accretion rate) can be plausibly explained by the gradual decrease in the super-Eddington outflow, leaving more material in g/s to be accreted.


\subsubsection{The origin of the UV/optical emission}

The source of early power-law UV/optical emission of TDEs is still uncertain. For instance, it may be produced by a reprocessing layer, which is powered by the X-ray emission emanating from the inner accretion disk \citep{Loeb+97, Guillochon+14, Metzger2016, Roth+16}, from the outer shock that forms at the intersection of the debris streams near their orbital apocenters \citep{Dai15,Shiokawa+15,Piran+15, BonnerotLu19,LuBonnerot19}, or from an elliptical accretion disk \citep{Zanazzi+20, Liu_2021}. 

The bolometric disk luminosity of {\tt Swift 1} can be inferred either from slim disk modeling ($1.4^{+4.0}_{-0.8}\times 10^{43}$ erg/s) or from the {\tt tdediscpec} \citep{Mummery+23} model $(2.2^{+5.3}_{-1.7}\times 10^{43}$ erg/s).  These models reach statistically consistent estimates, each of which is significantly fainter than the peak UV luminosity (see also Figure~\ref{fig:LRT}). If the UV/optical emission originates from a reprocessing layer, the disk would not be initially bright enough to power the layer (though at later epochs this conclusion no longer holds, see Figure \ref{fig:LRT}).
This indicates that the initial UV/optical emission in this event comes from shock dissipation or the outer regions of an elliptical accretion disk, and the lack of significant temperature or radius evolution suggests that this may remain the UV/optical power source at later epochs. In Section~\ref{CUX}, we will present additional evidence to support this conclusion.

Here, we show that the outer shock scenario can explain the brighter-than-the-disk UV/optical emission. Due to general relativistic pericenter precession, debris streams will undergo a self-intersection after each pericenter passage \citep{Rees1988,Dai15,Shiokawa+15, LuBonnerot19,BonnerotLu19, Andalman+22}.  Other shocks may also contribute to the hydrodynamic evolution of the debris \citep{Guillochon+14, Steinberg+22}, with the net result being dissipation of orbital energy and the formation of a disk with outer radius equal at least to the circularization radius $R_{\rm c}=2R_{\rm p}$, where $R_{\rm p}$ is the pericenter of the disrupted star. The rate of energy release through the circularization process is $\dot{E}_{\rm c}=\mu GM_{\bullet}\dot M/(2R_{\rm c})$ (here, we add $\mu\leq1$ to account for incomplete circularization, which would result in an elliptical rather than a circular disk). The circularization efficiency is then $\eta_c=\mu r_{\rm g}/(4R_{\rm p})$. For super-Eddington accretion disks, the disk radiation efficiency, $\eta_{\rm d}$, decreases significantly as the accretion rate increases, and typically $\eta_d<0.1$ for $a_\bullet<0.9$, due to advective cooling \citep{Abramowicz1988, Sadowski2009A}. For highly super-Eddington accretion disks, their bolometric luminosities increase over the Eddington limit as $\propto \ln \dot{m}$ \citep{Abramowicz2005}. As a result, the disk radiation efficiency $\eta_{\rm d}$ drops by a factor $\sim (\ln\dot{m})/\dot{m}$.  For the case of $M_\bullet = 2 \times 10^5 M_\odot$, we obtain $\eta_c = 0.002\mu \beta$ (where $\beta=R_{\rm p}/r_{\rm t}$ is the star's penetration parameter) and $\eta_d < 0.0005$ (assuming $\dot m \sim 1500$). For the case of $\mu=1$ and $\beta=1$, we find that $\eta_c > \eta_d$, resulting in the UV/optical bolometric luminosity being larger than the disk bolometric luminosity.

For the disruption of solar-like stars with $R_p = r_t$, the debris circularization efficiency ($\eta_c \propto M_\bullet^{2/3}$) decreases more slowly than the disk efficiency ($\eta_d \propto \sim M_\bullet^{3/2}$) as $M_\bullet$ decreases. The shock paradigm predicts that TDEs with $M_\bullet < \text{a few} \times 10^5~M_\odot$ will have a brighter peak UV/optical bolometric luminosity than the disk bolometric luminosity. Future observations of IMBH TDEs will help verify this prediction.

\subsubsection{BH mass inferred from UV/optical emission}
\label{UVMass}

Currently, various models are employed to fit UV/optical light curves 
and to estimate the black hole mass and other parameters of interest. Notable examples include {\sc MOSFiT} (Modular Open Source Fitter for Transients) \citep{Mockler19}, {\sc TDEmass} \citep{Ryu2020} and {\sc Redback} \citep{Sarin2023}. {\sc TDEmass} assumes that the UV emission originates from shocks near the apocenter of the most bound debris \citep{Piran+15}, while {\sc Redback} postulates that the UV emission arises from the cooling envelope formed by the returning debris \citep{Metzger2022}. {\sc MOSFit} makes fewer physical assumptions but instead posits a power-law relationship between various observables \citep{Guillochon+14}.  For our analysis, we utilize both {\sc TDEmass} and {\sc MOSFiT} to place constraints on TDE parameters.

{\sc TDEmass} posits that the dissipation of orbital energy and the production of UV/optical emission occur due to shocks near the debris apocenter. Its radiation efficiency is a function of the semimajor axis ($a_{\rm o}$) of the debris' orbit, $\eta_c\propto r_{\rm g}/a_{\rm o}$. Its temperature is determined by $T^4=L/(2\pi a_{\rm o}^2\sigma_{\rm sb})$. As $a_{\rm o}$ is a function of the specific energy of the debris, {\sc TDEmass} does not aim to fit the entire UV light curve but instead focuses on the peak emission, primarily constraining two key parameters: $M_\bullet$ and $M_\star$. The estimation of $M_\bullet$ relies heavily on the temperature observed at peak luminosity, while $M_\star$ primarily depends on the peak luminosity. We use the peak BB luminosity and temperature of the early epochs, $L=10^{44.28\pm0.75}$erg/s and $T=10^{4.72\pm0.17}$K, for the fit. 

{\sc MOSFiT} calculates the black-body UV/optical luminosity based on the mass fallback rate, assuming a time-independent efficiency ($\eta_{\rm c}$). The mass fallback rate $\dot M_{\rm f}(t)$, obtained from hydrodynamical simulations of TDEs \citep{Guillochon2013}, depends on parameters such as $M_{\bullet}$, $M_*$, and $\beta$. Instead of predicting the BB radius and BB temperature from a specific theoretical model, {\sc MOSFiT} fits these parameters to the observed data, assuming that the BB radius decreases as a power-law of the luminosity, $R_{\rm BB}\propto (L/L_{\rm Edd})^{l}= \dot m^l \propto t^{l'}$. This assumption aligns with predictions from certain theoretical models; for instance, elliptical disk models predict $R_{\rm BB}\propto t^{-1.1}$ \citep{Liu2014}, while cooling envelope models predict $R_{\rm BB}\propto t^{-1}$ \citep{Metzger2022}, although it is in disagreement with some radiation-hydrodynamics simulations \citep{Steinberg+22}. An additional assumption added in {\sc MOSFiT} is that the luminosity does not follow the mass fallback rate directly, but only follows convolution with a ``viscous delay,'' adding an additional free parameter to account for potential delays in the circularization process. 

{\sc MOSFiT} sets a soft cap to prevent the UV black-body luminosity from exceeding the Eddington luminosity. However, as we have shown, a super-Eddington UV luminosity is not impossible. As a result, we remove the soft cap from the code, allowing the luminosity to exceed the Eddington luminosity. As we have shown, the decay of the $w1$ magnitude as a function of time follows the mass fallback rate, a constant radiation efficiency may be correct for early UV emission of this event. The free parameters are $M_\bullet$, $M_\star$, $\beta$, viscous decay time $T_{\rm v}$, $\eta_{\rm c}$, BB radius normalization factor $n_{\rm R}$, $l$, and the first fallback time $t_0$ (days since first detection). We restricted our fitting to the $w2$, $m2$, and $w1$ light curves from the early epochs, and we applied priors for two specific parameters: $t_0$ and $\eta_c$, with their ranges set to [-30,0] days and [$10^{-4}$, 0.4], respectively. We use the Markov--Chain--Monte--Carlo routine in {\sc MOSFiT} to perform the fit and stop the calculation until it converged, with the potential scale reduction factor (PSRF) $\le1.1$ \citep{Gelman1992}.

Table \ref{tb:uv} shows our fit results.  Both {\sc TDEmass} and {\sc MOSFiT} produce consistent estimates for $M_\bullet$, which are also consistent with the result obtained from our X-ray spectral modeling (as illustrated in Figure~\ref{fig:maep4}).  In the case of {\sc TDEmass}, the determination of the stellar mass ($M_\star$) is primarily driven by the peak luminosity, $L_{\rm p}$ \citep{Ryu2020}. This reliance on $L_{\rm p}$ stems from the fact that {\sc TDEmass} fixes the radiation efficiency at $\eta_{\rm c}= r_{\rm g}/a_{\rm o}$, giving a relatively small value of 0.0003 (see Table \ref{tb:uv}). This value is considerably smaller than the circularization efficiency got by {\sc MOSFiT} ($\eta_c=r_g/(4R_p)=0.01$), which accounts for the significantly larger fitted values of $M_\star$ obtained by {\sc TDEmass} compared to those from {\sc MOSFiT}. The $R_{\rm BB}$ of {\sc MOSFiT} can be recovered from the relationship, $R_{\rm BB}=n_{\rm R} a_o (L/L_{\rm Edd})^l$. From the fit results, we see that $R_{\rm BB}$ evolves with time very slowly, $R_{\rm BB}\propto t^{-0.1}$, and $R_{\rm BB}\sim 0.7a_o$, consisting with the location of the emission  assumed by {\sc TDEmass} \footnote{Note that {\sc TDEmass} assume a solid angle of $2\pi$, while {\sc MOSFiT} assume a solid angle of $4\pi$. There should be a factor of $\sqrt{2}$ between the two $R_{\rm BB}$}. The {\sc MOSFiT} model fits the UV emission well, consistent with a stream self-interaction shock scenario. We note that the small fitted $M_\star$ value is consistent with the result from X-ray spectral fitting, which found between $\approx 0.005 - 0.065M_\odot$ of debris accreted during the early epochs. There are at least three possible reasons to account for the small fitted $M_\star$ and total mass accretion: 1) the disruption is attributed to the secondary of a binary system, leading to an orbital energy shift of the debris caused by the primary (for further analysis, we refer to Section~\ref{DBSVP}); 2) a partial disruption; 3) the super-Eddington UV luminosity blows out a significant amount of debris. In the first of these explanations, a shift in the orbital energy results in less debris being bound to the secondary, consequently leading to a smaller estimated value for $M_\star$.

\begin{table}
\centering
\caption{Best-fit parameters and their constraints derived from modelling the early UV emission with {\sc MOSFiT} and {\sc TDEmass}. We calculate the $\eta_c$ from the best-fit values of $M_\bullet$ and $M_\star$ for the case of {\sc TDEmass}. See Section~\ref{UVMass} for the meaning of the symbols.}
\begin{tabular}{ccc}
\hline
Parameter  & {\sc TDEmass} & {\sc MOSFiT}   \\
\hline
$M_{\bullet}$ / $10^5M_{\odot}$  & $5.4^{+10}_{-4.9}$& $8.8^{+10}_{-7.7}$   \\
$M_*$ / $M_{\odot}$ & $1.2_{-0.77}^{+88}$  & $0.07^{+0.12}_{-0.04}$  \\
$\eta_c/10^{-2}$ & $0.03^{+0.01}_{-0.02}$ & $1.0^{+3.0}_{-1.0}$     \\
$n_{\rm R}$  & - & $0.69^{+0.74}_{-0.21}$   \\
$\beta$  & -  & $0.9^{+0.01}_{-0.02}$  \\
$t_0$ / days & - & $-29^{+2}_{-1}$   \\
$T_{\rm v}$ / days & - & $0.03^{+0.32}_{-0.03}$   \\
$l$ & - &  $0.05^{+0.07}_{-0.04}$  \\
\hline
\end{tabular}
\label{tb:uv}
\end{table}

\subsection{X-ray versus UV/optical Timing}

In this section, we investigate whether there are any time lags between the X-ray and UV/optical emissions, to explore possible relationships or sources of the UV/optical and hard X-ray emission.

\subsubsection{X-ray Spectral Lags}
\citet{Zhang2022} reported tentative evidence of soft lags in one {\it XMM-Newton} observation (obs id: 0853980201; {\it XMM}~2) of AT2018fyk. In a soft lag the arrival time of lower energy photons lags behind that of higher energy photons. In their frequency-resolved analysis, the soft lags were detected in only the lowest frequency bin. To further investigate this result, we re-computed the frequency-resolved lags from this observation, using the same choice of energy bands and reference band (0.5--1 keV) as \citet{Zhang2022}. While finding results in general agreement with their analysis, the number of frequencies available in this lowest-frequency bin is small ($\sim$2), a factor of 2-3 less than recommended for reliable time lag estimates (\citealt{2014A&ARv..22...72U}). Given that the soft lags are detected in this bin only, and that the count rate of the observation is very low ($<$1.5 counts/s in the 0.3--5 keV band), we do not find the lags to be statistically significant. As an additional check, we computed the (non-frequency-resolved) lags using the linear Interpolated Cross-Correlation Function (ICCF) technique of \citet{Peterson_1998}, using the same energy bands/reference band as before. We again find no significant time lag, with the maximum correlation coefficient (often denoted $R_{\rm {max}}$) being less than 0.1 in all cases.

\subsubsection{Cross-Correlations between UV, Soft X-ray, and Hard X-ray Light Curves}
\label{CUX}

\begin{figure*}
    \centering
    \includegraphics[width=0.32\textwidth]{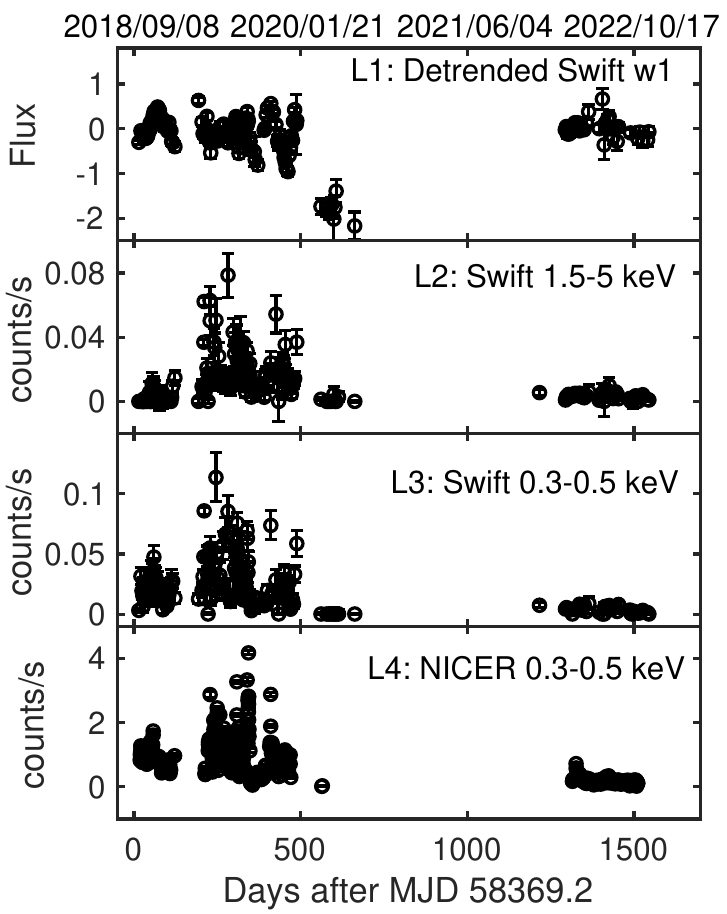}
    \includegraphics[width=0.313\textwidth]{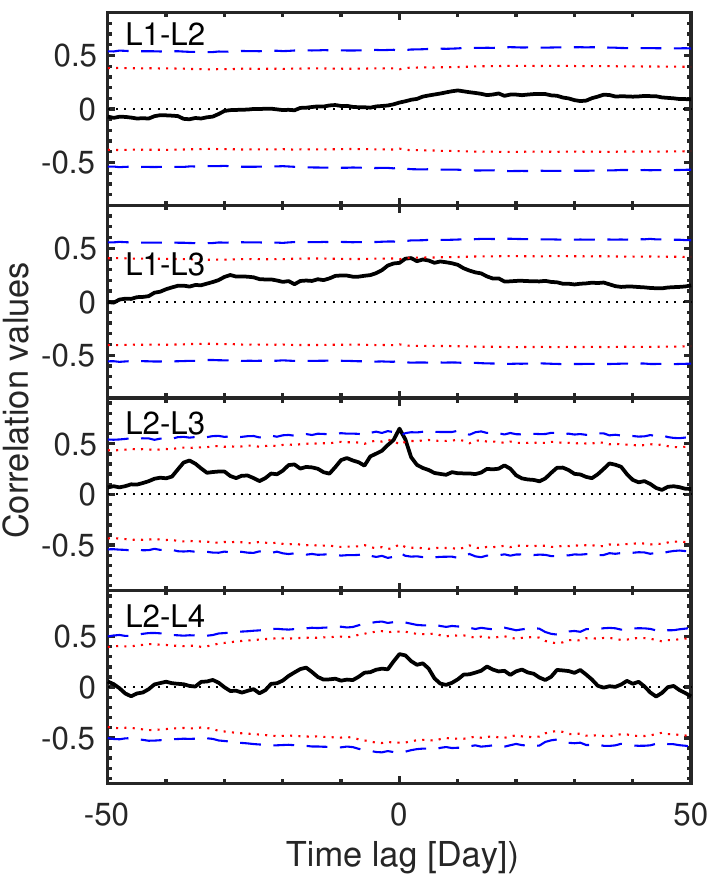}
    \includegraphics[width=0.32\textwidth]{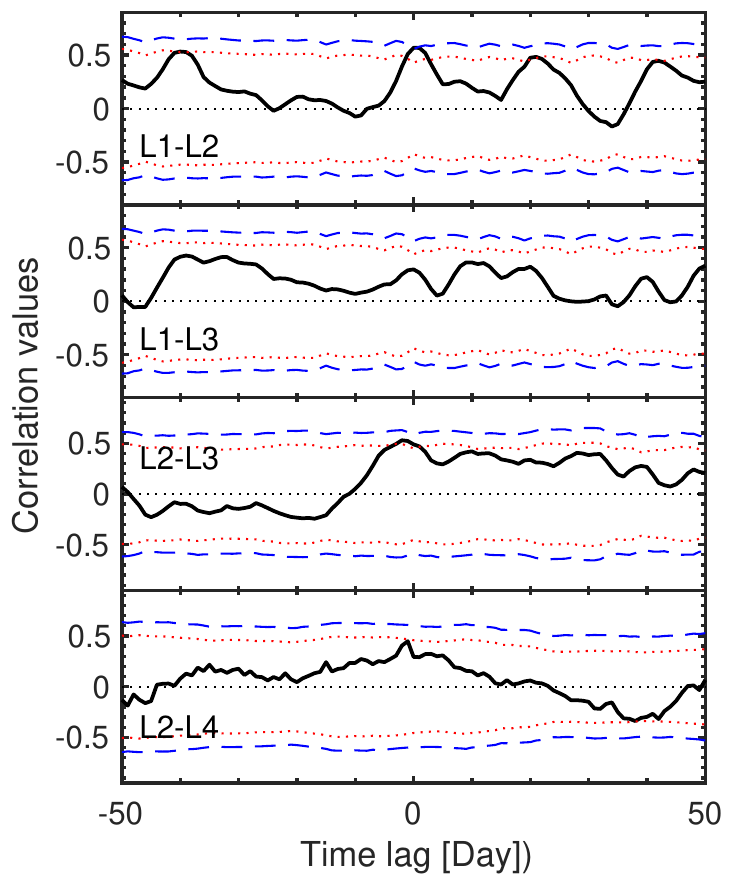}
    \caption{The light curves and the cross correlation functions. Left panels: The L1--L4 are as defined in the top right hand side of the left panels. From top to bottom; the {\it Swift} $w1$ (L1), 1.5--5 keV (L2), 0.3-0.5 keV (L3), and the NICER 0.3-0.5 keV (L4) light curves. 
    Middle and right panels: The black solid lines denote the correlation values of each time lag bin (with a 1-day width). The red dotted and the blue dashed lines show the $2\sigma$ and $3\sigma$ CLs, respectively. Positive time lag denotes the first light curve is leading. The middle panels show the correlation values of the early light curves, while the right panels show the correlation values of rebrightening light curves. These plots show marginal evidence for a correlation between the hard X-ray emission (L2) and the soft X-ray emission (L3 and L4). The centroids lags of L1 vs. L3, L2 vs.~L3 and L2 vs.~L4 of the middle plot are $3.8_{-1.9}^{+1.6}$, $-0.5_{-1.6}^{+0.9}$ and $1.4^{+19.3}_{-1.4}$ days, respectively.}
    \label{Lc}
\end{figure*}

We use the ICCF \citep{Peterson_1998} to explore whether there are correlations among the UV (L1; $w1$), soft X-ray (0.3-0.5 keV; L3 {\it Swift}; L4 NICER), and hard X-ray (1.5-5 keV; L2 {\it Swift}) emission. We use the package {\sc PYCCF} to calculate the ICCF \citep{Sun2018}. We estimated the uncertainty in the peak and the centroid of the ICCFs by using the Random Subset Selection and the Flux Randomization procedures \citep{Peterson_1998,Peterson_2004}.The study \citep{Welsh1999} shows that extended trends in light curves can introduce biases in the ICCF analysis. Consequently, removing these trends by fitting the light curves with a smooth function enhances the accuracy of recovering correlations and lags between two considered time series. The {\it Swift} w1 light curve exhibits a power law decay during the initial $\sim500$ days and subsequent rebrightening after $>1200$ days. We detrend the w1 light curve by subtracting the best-fit power law values, as detailed in Section \ref{uvothst}. However, for the X-ray light curves, there are no clearly discernible long-term trends, as depicted in the left panel of Figure \ref{Lc}. Consequently, we opt not to detrend them during the ICCF analysis. We calculate the ICCF of the early light curves and the rebrightening light curves separately. In order to determine the confidence of the correlations, we generate $10^5$ random white noise X-ray light curves using the method described in \cite{Timmer1995}.  The randomly generated X-ray light curves require a power-law index, which relate power and frequency. We adopt the power-law index as $1.49$ and $1.32\pm0.3$ for the band 0.3-0.5 keV and 1.5-5 keV \citep{Zhang2022}, respectively \footnote{Our conclusions are unchanged for a different index of 0.5 or 2.5 for both bands.}. We then cross-correlate the light curves with each of these simulated X-ray light curves, to built a distribution of $10^5$ correlation values at each time lag bin. Using these distributions, we extracted the $95\%$ ($2\sigma$) and $99.7\%$ ($3\sigma$) confidence contours. 

Figure~\ref{Lc} show the light curves and the correlation values of each time lag bin.  The time lags of L1 vs.~L2 and L1 vs.~L3 are $17^{+23}_{-8}$ days and $3.8^{+1.6}_{-1.9}$ days for the early emission, respectively. However, the significance of the two time lags is less than $2\sigma$. During the rebrightening phase, the positive lag between L1 and L3 weakens, and four high correlation values appear for L1 and L2. This may be attributed to the bad observed cadence and low detected X-ray count rates being suboptimal. Recently, \cite{Guo2023} found a similar lag of $9.2^{+6.1}_{-6.0}$ between the w1 and X-ray emission, limited to the initial 120 days. The positive time lag between L1 and L3 here aligns with that from \cite{Guo2023} within $1\sigma$ CL. As proposed by \cite{Guo2023}, the positive lag disfavors the assumption that the early UV/optical emission originates from a reprocessing layer but supports the paradigm that the early UV/optical emission comes from the outer shocks. Nevertheless, the lack of a significant correlation between the UV and X-ray bands seems consistent with the X-ray source being centrally compact and dominated by light bending close to the SMBH. These findings are in line with the early analysis suggesting that the UV/optical emission originates from the outer shocks.

The correlation analysis between hard and soft X-ray emissions reveals a significance above the $3\sigma$ confidence level for the L2 vs. L3 case during the initial 600 days. The calculated time lag is $-0.5^{+0.9}_{-1.6}$ days, consistent with a lag of 0 days. This 0-time lag is also observed in the L2 vs. L3 case during the rebrightening phase and the L2 vs. L4 case in both the early and rebrightening phases. However, it is noteworthy that the significance of the correlation diminishes in the rebrightening phase and between L2 and L4, possibly due to poor cadence during that period and variations in the {\it NICER} observation cadence. Despite this, the persistent 0-time lag suggests a correlation between the hard and soft components. This finding aligns with our analysis in Section~\ref{hardXray}, supporting the hypothesis that the hard component originates from the corona above the disk, which generates the observed soft X-ray emission. A correlation between the hard and soft X-ray emission is also shown in some AGN system, e.g. Narrow-line Seyfert 1 galaxy 1H 0707-495 \citep{Robertson15}, IRAS 13224-3809 \citep{Chiang15}. We also conducted the ICCF by de-trending the X-ray light curves through subtracting the best-fit values of the power-law model. The results also align with the findings here, e.g. no significant correlation between $w1$ and soft- or hard X-ray emissions, and a $3\sigma$ confidence level correlation between soft and hard X-ray emissions for case of L2 vs. L3 in the initial 600 days.

\section{Discussion: A TDE in a Binary SMBH system?} \label{sec:binary}
In this section, we interpret our previous data analysis as evidence for the production of a TDE in a binary SMBH system and explore its implications.

\subsection{Discrepancy with Black Hole Mass from Galaxy Scaling Relation}

The lack of a high equivalent width H-$\delta$ absorption line implies that the galaxy does not belong to the E + A galaxy class \citep{Wevers_2019}. \citet{Wevers20} measured the velocity dispersion of the host galaxy of the AT2018fyk to be $\sigma_*=159\pm1 \rm{km~ s^{-1}}$. Using the M--$\sigma_*$ relation of \citet{Mcconnell2013}, \citet{Wevers20} estimated the host BH mass as $M_\bullet=10^{ 7.7\pm0.4} M_{\odot}$. Adopting the M--$\sigma_*$ relation of \citet{Ferrarese2005}, the host BH mass is $M_\bullet=10^{ 7.7\pm0.4} M_{\odot}$, whereas it is $M_\bullet=10^{ 8.0\pm0.4} M_{\odot}$ for the M--$\sigma_*$ relation of \citet{Kormendy2013}. A measurement from the $M_\bullet$--$M_{\rm bulge}$ relation also supports the inference of a high-mass SMBH, for example, \cite{Wevers_2019} found that $M_\bullet=10^{7.3\pm0.4}M_\odot$ using the $M_\bullet$--$M_{\rm bulge}$ relation of \cite{Haring2004}.

In contrast, the $M_\bullet$ inferred from our fiducial X-ray spectral fits is orders of magnitude lower, $(2.7^{+0.5}_{-1.5})\times10^5M_\odot$. Even considering a (non-fiducial) joint fit with all the epochs, the X-ray parameter estimation still rules out $M_\bullet>7\times10^5M_\odot$ at $>3\sigma$ CL. Likewise, the $M_\bullet$ inferred from the UV emission is $10^{4.7}-10^{6.3}M_\odot$, further supporting the small SMBH mass inferred from the X-ray spectral fits. The discrepancy between the inferred $M_\bullet$ from X-ray spectral fits and that derived from the M-$\sigma$ relation is consistent with the idea that the thermal X-ray flare after the disruption is not generated from a disk around the host galaxy's primary SMBH, but rather from a disk around a BH with much smaller $M_\bullet$, orbiting close to the central SMBH.

Arguments based on galaxy scaling relations are of course approximate.  The M--$\sigma_*$ measurement may over estimate the inferred $M_\bullet$, e.g. \cite{Koss2022} found that the M--$\sigma_*$ relation of \citet{Kormendy2013} overestimates $M_\bullet$ by 0.7 dex (this would be 0.4 dex if using the relationship of \citealt{Ferrarese2005} and \citealt{Mcconnell2013}) by comparing to the direct $M_\bullet$ measurement from megamasers ($M_\bullet<10^{8.5}M_\odot$). Such an overestimate is also reported by \cite{Greene2016}. In general, however, even these possible systematic biases are unlikely to resolve the $\sim2.5$ order of magnitude discrepancy between X-ray continuum fitting measurements of $M_\bullet$ and estimates from galaxy scaling relations.

\subsection{Rapid UV Light Curve Decline Suggesting Accretion Cutoff}
\label{BBH}

A star with mass $M_*$ and radius $R_*$ will be tidally disrupted as it approaches a SMBH within the tidal radius, $r_t=R_*(M_\bullet/M_*)^{1/3}$ \citep{Rees1988, Phinney1989}. After the disruption at the tidal radius, the specific energy E across the stellar radius varies from $E_o-\Delta E$ to $E_o+\Delta E$, where $E_o$ is the original orbital binding energy of the star, and $\Delta E= GM_\bullet R_*/r_t^2$ is the spread in specific energy across the stellar radius. Typically, $E_o$ is negligible compared to $\Delta E$, as most stars tidally disrupted in realistic galactic nuclei approach the SMBH on nearly zero energy orbits whose apocenters lie at parsec scales \citep{Magorrian1999,Wang&Merritt2004}. The bound debris with $E<0$ moves in high eccentricity Keplerian orbits, and falls back to the tidal radius after one Keplerian period, $T=2\pi G M_\bullet(-2E)^{-3/2}$. The debris returning to the tidal radius generates a mass fallback rate,
\begin{equation}
\dot M_{\rm f}=133 \dot M_{\rm Edd} M_6^{-3/2}m_*^2 r_*^{-3/2} \left(\frac{t}{t_{\rm f}}+1\right)^{-5/3},
\end{equation}
where $m_\star=M_\star/M_\odot$ and $r_\star=R_\star/R_\odot$. The return time of the most bound debris is
\begin{equation}
\label{tfall}
    t_{\rm f}= 40  {~\rm days} \times M_6^{1/2}m_*^{-1}r_*^{3/2}.
\end{equation}
As the (initial) viscous time scale is much shorter than this orbital period, debris will be accreted quickly once it returns to pericenter, so long as circularization is effective (e.g. \citealt{Steinberg+22}, though see also \citealt{Ryu+23}).  If this is the case, an accretion disk will be formed and its accretion rate will initially follow the mass fallback rate, deviating at later times once the viscous time grows \citep{Cannizzo+90, vanVelzen+19}.

In a SMBH binary system, the star can in principle be disrupted by both the primary, of mass $M_{\bullet, \rm p}$, and the secondary, of mass $M_{\bullet, \rm s}$ \citep{Liu_2009,Coughlin2017}. After disruption, the most tightly bound debris falls back to the pericenter of the SMBH that caused the TDE as in a single SMBH system; gravitational torques from the companion SMBH cause only small perturbations on the slow secular timescale \citep{Kozai1962, Lidov1962, Liu2014}. For less tightly bound debris, with an apocentre larger than a critical radius, fluid elements will move on chaotic orbits \citep{Liu_2009,Coughlin2017} due to the lack of a hierarchy in the gravitational fields of the two SMBHs.  As a result, this gas will fail to fall back to the pericenter after disruption, producing an accretion cutoff at a time $T_{\rm c}$ \citep{Liu_2009}. Following \cite{Coughlin2017}, we take the Hill sphere as this critical radius. As we have shown that the TDE flare is due to the smaller BH, the corresponding Keplerian time $T_{\rm c}$ can be estimated as \citep{Coughlin2018}
\begin{equation}
\label{Tc}
T_{\rm c}= 22 {~\rm day}\times M_{\rm p,8}^{-1/2}\left(\frac{a}{10^{-3}{\rm pc}}\right)^{3/2},
\end{equation}
here, $M_{\rm p,8}=M_{\bullet,\rm p}/(10^8M_\odot)$ and $a$ is the separation of the binary. 

The cutoff time $T_{\rm c}$ can be directly measured by observations, and used to constrain a combination of $a$ and $M_{\rm p}$.  AT2018fyk was discovered on 2018 September 8 (MJD 58369.23), with a non-detection reported in the $g^\prime$ band ($g^\prime>$ 17.4 mag) on 2018 August 29 (MJD 58359). As shown in Figures~\ref{fig:UV} \& \ref{Lc}, there is a rapid decline for both UV/optical and X-ray light curves on 2020 March 22 (MJD 58930.15), indicating an accretion cutoff starts approximately on this day. The last observation before the sharp cutoff is on 2020 Jan 09 (MJD 58857.6). The upper and lower limit of $T_{\rm c}$ are therefore 571 and 488 days, respectively. When the disk is cut off from an external resupply of debris, the disk accretion rate 
will drop quickly, over a timescale much smaller\footnote{More specifically, the accretion rate in an isolated disk evolves as $\dot M\propto (t_{\rm v}/t)^{4/3}$ \citep{Cannizzo+90, Liu2014}. Here, the viscous time $t_{\rm v}\approx0.034~{\rm days} \times \alpha^{-1}\beta^{-3/2}(H/R_c)^{-2}$. 
 The initial ($H/R \sim 0.5$) viscous time at $R_{\rm c}$ is thus $t_{\rm v} \sim 1 ~\rm{days}$.  As can be seen from Figure \ref{fig:LRT}, the BB luminosity drop by a factor of $\sim10$ during the long declining phase, but if the disk were isolated, the time for the accretion rate to drop by a factor of 10 would be a far shorter $5.6t_v$ $\sim 8$ days, showing that the initial evolution is fallback-regulated rather than viscously regulated.} than the uncertainty on $T_{\rm c}$, and we therefore neglect this complication. We thus take $T_{\rm c}=530\pm40$ days. Together with $M_{\bullet, \rm p}=10^{7.7\pm0.4}M_\odot$, we obtain $a=(6.7\pm1.2) \times 10^{-3}{~\rm pc}$. 

Assuming a circular orbit, the period of this binary is $T_b=2\pi a^{3/2}/\sqrt{GM_{\bullet, \rm p}}=7.4\pm2.2 {\rm ~yr}$. When we use a different accretion cutoff assumption, as suggested by \cite{Liu_2009}, we obtain a similar binary period, $T_b\approx T_c/\epsilon=4.8\pm2.7$ yr, with semi-empirical constant $\epsilon = 0.32 \pm 0.18$. This result is consistent with the $7.4\pm2.2 {\rm ~yr}$ above. The merger time of a binary can be estimated as \citep{Peters1964}
\begin{align}
    T_{\rm m}=&580 {\rm ~yr}\times M_{p,8}^{-3}\left(\frac{a}{10^{-3}\rm pc} \right)^4\frac{M_{\bullet, \rm p}^2}{M_{\bullet, \rm s}(M_{\bullet, \rm p}+M_{\bullet, \rm s})} \notag \\
    \approx&1.8 \times 10^9~{\rm yr}~  \left(\frac{T_{\rm c}}{530~\rm d} \right)^{8/3} \left(\frac{M_{\bullet, \rm p}}{10^{7.7}M_\sun} \right)^{-2/3} \\
    & \times \left(\frac{M_{\bullet, \rm s}}{10^{5.4}M_\sun} \right)^{-1}    . \notag
\end{align}
In the second equation, we have approximated $M_{\bullet, \rm p} \gg M_{\bullet, \rm s}$. 
 With the values of $M_{\bullet, \rm p}$, $M_{\bullet, \rm s}$ and $T_{\rm c}$ as given above, we obtain $T_{\rm m}=1.8^{+1.8}_{-0.9}\times 10^9 {\rm yr}$.  This result passes a non-trivial consistency test for our model, which would have suffered from fine-tuning if we had determined $T_{\rm m} \ll$ the typical time between SMBHB formation (e.g. the galaxy merger timescale).

\subsection{Debris Bound to Secondary versus Primary}
\label{DBSVP}

Here, we explore the consequences of the disruption being done by the secondary SMBH in the binary. We define an effective energy, $E_{\rm eff}$, instead of the energy $E$, to take the impact from the presence of the primary SMBH into account, $E_{\rm eff}=V_{\rm E}^2/2-GM_{\bullet, \rm s}/r_{s,t}$,
where $V_{\rm E}$ is the debris velocity relative to the secondary at the tidal radius, $r_{s,t}$ is the tidal radius of the secondary. $V_{\rm E}$ can be determined by,
$V_{\rm E}^2/2=V^2/2+V_o^2/2+VV_o\cos\phi$. Here, $V\approx\sqrt{ 2E+2GM_{\bullet, \rm s}/r_{s,t}+2GM_{\bullet, \rm p}/a}$ is the total velocity of the debris, $V_o=\sqrt{GM_{\bullet, \rm p}/a}$ is the orbital velocity of the secondary and $\phi$ is the difference in the direction between $V$ and $V_o$. Setting $f=2V\cos \phi/V_o+3$ (typically, $V>V_o$), we get $V_{\rm E}^2/2= E+GM_{\bullet, \rm s}/r_{s,t}+fGM_{\bullet, \rm p}/2a$.
Following \citet{Coughlin2017}, we define an energy ratio,
\begin{equation}
K_E=\frac{GM_sR_*}{r_{s,t}^2}\frac{2a}{GM_{\bullet, \rm p}}.
\end{equation}
The effective energy of the most bound debris to the secondary can be written as, 
\begin{equation}
E_{\rm eff}\approx -\Delta E+\frac{fGM_{\bullet, \rm p}}{2a}=(f/K_E-1)\Delta E.
\label{eff}
\end{equation}
If $f>K_E$, no debris will return to the pericenter of secondary (this is only roughly correct, as debris that can go through the Roche lobe of the secondary can also approach the secondary again). After the disruption, the bound debris with effective energy $E_{\rm eff}<0$ (and an apocenter less than the Hill sphere) moves in approximately Keplerian orbits of the secondary. Due to the impact from the primary, the fallback time scale and peak accretion rate are modified by a factor of $(1-f/K_E)^{-3/2}$ and $(1-f/K_E)^{5/2}$, respectively, compared with the disruption by single BH. We note that, this analysis is consistent with the simulation of \citet{Coughlin2018}, who constructed 30 disruptions subjected to the secondary, where 23 yield no accretion to the secondary, and 2 cases show fall back later or earlier than the single BH disruption case. 

If the debris that goes beyond the Hill sphere of secondary can be accreted by primary, the most bound debris to the primary should have an energy of $E\approx-(f+(9q)^{2/3})\Delta E/K_E=-\kappa\Delta E$ (note that $\Delta E=GM_{\bullet, \rm s}R_*/r_{s,t}^2$, $\kappa=f/K_E+(9q)^{2/3}/K_E$ and $q=M_{\bullet, \rm s}/M_{\bullet, \rm p}$). We assume the debris has been accreted after it orbits the primary by one Keplerian orbit. The fallback rate to the primary can be approximated as,
\begin{equation}
\dot M_p= \frac{M_*\kappa}{3t_{\rm f}}\left(\frac{t+t_{\rm f}}{t_{\rm f}}\right)^{-5/3}=\kappa^{5/2}q^{1/2}\dot M_{\rm p, f}.
\end{equation}
where $t_{\rm f}$ is the returning time of the most bound debris to the primary, and
$t_{\rm f}=t_{\rm s, f}\kappa^{-3/2}q^{-1}$ ($t_{\rm s, f}$ is the fallback of the most bound debris to the secondary, defined as in Eq.~\ref{tfall}). Here, we estimate the peak accretion rate to the primary for case of $\kappa=1$ (indicating no accretion to the secondary) and $M_\star=M_\odot$. We get $\dot m_p\approx0.02$ in Eddington mass accretion rate units and $t_{\rm f}\approx5000$ days for case of $M_\bullet=10^{7.7}M_\odot$. This accretion rate results in a bolometric luminosity of $L_{\rm d}\approx1.4\times 10^{44}$ ~erg/s, which is brighter than the rebrightening UV black body luminosity $L_{\rm BB}\approx10^{43}$erg/s, assuming a radiative efficiency of $\eta=0.1$. Consequently, it is possible that the rebrightening of the UV/optical emission originates from accretion onto the primary. However, we have also demonstrated that the rebrightened UV/optical emission closely follows the mass fallback law of $t^{-5/3}$. If this UV/optical emission were primarily attributed to accretion onto the primary, the fallback timescale should align with the rebrightening date, at approximately 1200 days. Nevertheless, the predict fallback timescale is significantly longer than this. 
Based on the aforementioned analysis, we favor an explanation in which the rebrightening epochs are a consequence of accretion onto the secondary. The UV/optical brightness during the early stages of the rebrightening epochs surpasses that observed near the cutoff at approximately $\sim 500$ days. This can be explain as, the debris's angular momentum, relative to the secondary, starts to differ from the material initially comprising the disk, resulting in an higher accretion rate \citep{Coughlin2017}. For TDE subjected to the secondary in a binary system with a small $q$, simulations \citep{Coughlin2018} indicate that, in general, there is negligible accretion onto the primary. If any accretion does occur, it typically happens at very late times, and the rate is very low. The associated timescale for the debris traveling directly to the primary falls in the range of $T_{\rm acc} = a/v_{\rm escape} \approx 6$ years for this event. 

By using Jacobi's constant \citep{Roth1952}, which describes the total energy of the debris, we can constrain $\kappa$ by estimating the mass accreted by the secondary before the first light curves cutoff. As the star is disrupted by the secondary at $r_{s,t}$, debris with Jacobi's constant bigger than the Jacobi's constant at the Lagrange points L1 would not be able to travel through the Roche lobe of the secondary. For small $q$, the Roche lobe can be approximated as the Hill sphere radius, $R_H=(q/3)^{1/3}a$, and Jacobi's constant at L1 can be written in the form, 
\begin{equation}
C_{\rm L1}\approx\frac{GM_{\bullet, \rm p}}{a}(1-R_H/a)^2+\frac{2GM_{\bullet, \rm p}}{a-R_H}+\frac{2GM_{\bullet, \rm s}}{R_H}. 
\end{equation}
The Jacobi's constant of debris with a different energy E at the disrupted position can be estimated as,
\begin{equation}
C_{\rm E}\approx\frac{GM_{\bullet, \rm p}}{a}(1-r_{s,t}/a)^2+\frac{2GM_{\bullet, \rm p}}{a-r_{s,t}}+\frac{2GM_{\bullet, \rm s}}{r_{s,t}}-V_{\rm E}^2.
\end{equation}
We assume that debris with $C_E>C_{\rm L1}$ is all accreted before the first light curves cutoff. As a result, the mass ratio between the debris accreted by secondary and the mass of the star is,
\begin{equation}
    f_m=\frac{C_E-C_{\rm L1}}{4\Delta E} \approx\frac{1-\kappa}{2}.
\end{equation}
$f_m$ and $\kappa$ can be constrained from observation. By fitting the X-ray spectra, we get that the mass accreted by secondary is $\approx0.005$---$0.065 M_\odot$ at $1\sigma$ CL, during the first 561 days. Note that this mass is in good agreement with the $M_\star=0.07^{+0.12}_{-0.04}M_\odot$ inferred from the MOSFiT light curve fits. Assuming a solar mass star, we get $f_m$ is about 0.01--0.13 and $\kappa$ is about 0.984--0.74. Taking $\kappa=0.74$, the values of $\dot m_{\rm p}$ and $t_{\rm f}$ above change from 0.02 to 0.01 and from 5000 days to 7900 days, respectively.


\subsection{Binary TDE Event Rate}

An isolated BH in a galactic nucleus will tidally disrupt stars due to diffusion in orbital angular momentum space \citep{FrankRees76}.
In this picture, stars undergo a random walk in angular momentum from uncorrelated two-body scatterings, eventually reaching pericenters less than the tidal disruption radius. The outcomes of this random walk can be quantified in terms of an angular momentum ``loss cone,'' which is emptied on a short dynamical timescale and refilled on a slow relaxational timescale.  The TDE rate in such a system is thus equal to the loss-cone feeding rate produced by the two-body relaxation process \citep{LightmanShapiro77, CohnKulsrud78}.  Theoretical loss cone calculations generally find average TDE rates $\sim {\rm few} \times 10^{-5} - {\rm few} \times 10^{-4}~{\rm yr}^{-1}$ \citep{Magorrian1999,Wang&Merritt2004,Stone2016, Pfister+20, Stone+20,Teboul2024}, in some tension with (lower) recent TDE rate estimates \citep{Yao+23}.

The formation and evolution of a bound SMBHB in a galaxy's centers would significantly change TDE event rates. Shortly after the formation of a hard SMBHB, interactions between the binary and a bound stellar cusp will enhance the TDE rate to as high as 1 yr$^{-1}$, through a combination of the EKL mechanism \citep{Kozai1962,Lidov1962,Ivanov20 05} and strong three-body scatterings \citep{Chen2011,Wegg2011}. \cite{Mockler2023} also recently showed that around the smaller black hole, a modest binary eccentricity will allow the EKL mechanism \citep{Ford+00, Naoz2016} to produce an even larger number of TDEs from the smaller secondary BH.

The combined effect of secular and strong interactions between the binary and the surrounding dense stellar cusp will produce a number $N_{\rm b}\approx 7\times 10^4 q^{(2-\gamma)/(6-2\gamma)}M_7^{11/12}$ TDEs \citep{Chen2011}, where $1.5\leq\gamma \leq2$ is density profile index. The $N_{\rm b}$ estimation of \cite{Chen2011} is in general agreement with the simulations of \cite{Coughlin2017}, who found $N_{\rm b}\approx800-8000$ for case of $q=0.2$ and $M_{\bullet, \rm p}=10^6M_\odot$ on circular SMBHBs. The probability of stellar disruption by the secondary is about a factor of $q/2$ to that of primary in circular SMBHBs \citep{Coughlin2017}.  Supposing that a galaxy on average experiences N SMBHB episodes during a Hubble time ($10^{10}$ years), and assuming an observationally motivated TDE rate of $7.4\times 10^{-5}~{\rm yr}^{-1}$ \citep{Yao+23}, then during one duty cycle of galaxy merger, the relative numbers of TDEs by single BHs ($N_{\rm s}$), SMBHB primaries ($N_{\rm b}$), and SMBHB secondaries ($N_{\rm bs}$) are $N_{\rm s}:N_{\rm b}:N_{\rm bs}\approx21:2q^{0.1}M_7^{11/12}N:q^{1.1}M_7^{11/12}N$ (assuming $\gamma=1.75$). For the case of AT2018fyk, $q\approx0.004$ and $M_{\bullet, \rm p}\approx 10^{7.7}M_\odot$, gives $N_{\rm s}:N_{\rm b}:N_{\rm bs}\approx2100:500N:N$. It is worth noting that this estimation introduces some uncertainty due to the following factors: 1) Isolated SMBHs having higher TDE rates, leading to a lower binary TDE rate; and 2) the EKL mechanism, resulting in more TDEs around the secondary \citep{Mockler2023}, thereby contributing to a higher binary TDE rate due to the secondary.

The TDE Catalog currently has 116 TDEs candidates \citep{Goldtooth2023}.
If we conservatively take $N=1$, there would be probabilities of $\approx 100\%$ and $\approx 4\%$ to detect at least 1 TDE from the primary and the secondary, respectively, of an SMBHB out of a sample of this size. EKL mechanism can produce TDEs around the secondary that are comparable to those around the primary in the case of q=0.1 \citep{Mockler2023}. Assuming that the EKL mechanism can increase the TDE rate by an order of magnitude for other q values, the probability of detecting 1 TDE caused by the secondary out of 116 samples would then increase from $4\%$ to $31\%$. Currently, there are two TDEs, SDSS J120136.02+300305.5  \citep{Liu2014} and OGLE16aaa \citep{Shu2020}, explained as SMBHB TDE originating from the primary. The low detected rate of binary TDEs may be due to the fact that, it is not easy to distinguish the binary TDE from the single SMBH TDE, as the perturbation from the secondary is not strong enough to produce a significant difference in the light curves. However, some factors may make the observed rated of binary TDE due to secondary higher than predicted, e.g. the direct capture of the star by the primary SMBH with  $M_\bullet>10^7 M_\odot$ due to relativistic effects, would result in a substantial reduction in the rate of TDEs by the primaries in SMBHBs \citep{Coughlin2022}.



\subsection{Distinguishing between SMBHB TDE and repeating pTDE scenarios}

The rebrightening of this event has also been explained as a repeating pTDE with $M_\bullet=10^{7.7} M_\odot$ \citep{Wevers_2023}. In this scenario, a star, on a $\sim$1200-day orbit around the SMBH, is periodically stripped of mass during each pericenter passage, powering its unique light curves. As shown above, the binary TDE hypothesis can also explain the observations well and presents certain advantages over the repeating pTDE scenario. These include: (i) the detection of a BH mass, through multiple distinct methods, that is two orders of magnitude below predictions from host galaxy scaling relations; (ii) the large-amplitude fluctuations and sharp cutoff in the light curves; and (iii) the alignment of late-time mass fallback behavior with the $t^{-5/3}$ law rather than the $t^{-9/4}$ law expected for pTDEs \citep{Coughlin2019b}.  However, none of these arguments decisively rules out the possibility of a repeating pTDE. Galaxy scaling relations possess intrinsic scatter \citep{Kormendy2013} and major outliers do exist \citep{vandenBosch+12}, 
a repeating pTDE model can still explain the light curve cutoff \citep{Wevers_2023}, and the light curve fluctuations make that the light curve decay could be consistent with the $t^{-9/4}$ law (see Fig. \ref{fig:UV}). Here, we list possible (future) observations that can distinguish between these two scenarios, either for this TDE or for others: 
\begin{enumerate}
\item 
The time interval of the repeating rebrightening: Binary TDEs predict chaotic rebrightening times, while the pTDE predicts the same rebrightening time interval.
\item 
The blue/redshift of the emission lines \citep{Gaskell1983,DeRosa2019}: the binary TDE model predicts periodic blue/red-shift evolution of the emission lines due to the Doppler effect when the secondary moves with respect to our line of  sight, while the pTDE does not.
\item 
The ultimate fading of UV/optical emission: the binary TDE model predicts that the emission will drop to the host level in about 5000 days, whereas pTDE predicts that the flare would die out once the core stops being disrupted, which can be on either much shorter or much longer timescales than 5000 days.
\item 
Future gravitational wave detections with pulsar timing arrays (PTAs; e.g. \citealt{Agazie2023, Antoniadis2023}) may help resolve SMBHB hosts of TDEs with unusual light curve features, but this will only be possible for nearby sources, high-mass primaries, and secondaries that are as high-mass as possible.  While the chirp mass that we infer for AT2018fyk is too low to ever be detected as a resolvable PTA source \citep{Sesana+09}, a binary with $M_{\bullet, \rm p} \sim 10^9 M_\odot$ and $M_{\bullet, \rm s} \sim 10^7 M_\odot$ would be more promising.

\end{enumerate}

\section{Conclusions} \label{sec:Conclusions}

We conducted a comprehensive analysis of AT2018fyk, scrutinizing both its X-ray and UV/optical emission. For the X-ray spectra, we employed a model comprising a thermal disk and a thermally comptonized continuum. This allowed us to analyse the source of the soft and hard X-ray spectral component and place constraints on parameters such as $M_\bullet$ and $a_\bullet$ through X-ray spectral fitting. To understand the temporal behavior of the source, we fitted the $w1$ light curve to determine its decay rate. We examined the UV/optical light curves to investigate the evolution of the black-body temperature, radius, and luminosity. We investigate the origins of the early-time UV/optical emission and derived $M_\bullet$ from these data. Next, we studied the cross-correlation between the X-ray, UV/optical, and soft and hard X-ray emission to uncover potential connections and emission mechanisms. Finally, our interpretation of AT2018fyk centers on it being a supermassive binary TDE. We placed constraints on the binary parameters and also list possible observations that can distinguish between the binary TDE scenario and the repeating pTDE scenario. Our findings are:
\begin{enumerate}

\item 
The fit to the X-ray spectra shows that the soft X-ray spectrum can be explained as a disk spectrum, while the hard X-ray spectrum can be explained as emission from a corona above the disk.

\item
The $M_{\bullet, \rm s}$ and $a_\bullet$ are constrained to be $M_{\bullet, \rm s}=(2.7^{+0.5}_{-1.5})\times10^5M_\odot$ and $a_\bullet>0.3$ at the $1\sigma$ CL from slim disk modeling.

\item
Both the early and rebrightened $w1$ light curves decays as a power-law, following the theoretical mass fallback rate. The $M_{\bullet, \rm s}$ is estimated to be $(5.4^{+10}_{-4.9}) \times10^5 M_\odot$ and $(8.8^{+10}_{-7.7}) \times10^5 M_\odot$, as determined from the early time UV emission using the {\sc TDEmass} and {\sc MOSFiT} light curve modeling tools, respectively. These values are consistent with that inferred from our X-ray spectral modeling. 

\item 
The constant fitted BB temperature, slowly decreasing BB radius, and the higher BB luminosity compared to the expected disk luminosity suggest that the UV/optical emission originates from the circulation of debris streams. Our analysis shows that the early BB luminosity has super-Eddington values.

\item
The evidence for a correlation between the soft and hard X-ray emission supports the conclusion that the hard X-ray emission comes from the corona,  and the lack of significant correlation between the X-ray and UV/optical emission support the idea that the UV/optical emission originates from the circulation of the debris, whereas the X-ray emission comes from the inner accretion disk. 

\item
If the star is disrupted by the secondary of a binary SMBH, the rapid decline in the light curve occurring approximately 561 days after its discovery can be explained as the outcome of accretion onto the secondary experiencing a cutoff. Utilizing this specific point in time, in conjunction with $M_{\bullet, \rm s}$ and $M_{\bullet, \rm p}$, we constrain the separation of this binary to be $(6.7\pm1.2) \times 10^{-3}$ pc, and the corresponding merger time to be $1.8^{+1.8}_{-0.9}\times 10^9 $ years.

\end{enumerate}

The behavior of X-ray and UV/optical emission is consistent with that of IMBH TDEs, such as a high disk temperature, long super-Eddington time \citep{Lin2017, Lin2018, Wen_2021, Cao2023}, long UV power-law decay time, and higher early UV/optical bolometric luminosity than the disk bolometric luminosity. The $M_\bullet$ measured both from modeling the X-ray emission and independently from modeling the UV emission, strongly supports that this is an IMBH TDE. Notably, the $M_\bullet$ derived from TDE characteristics appears to be incommensurate with that determined through the $M_\bullet-\sigma_\star$ relation. We posit that this incongruity stems from the disruption of a star by the secondary of a SMBHB. Considering the protracted mass fallback time required for accretion onto the primary, we attribute the rebrightening phase to accretion onto the secondary. Consequently, we propose that both the early and rebrightening phases of radiation emission can be ascribed to the accretion process onto the secondary. The strong perturbation from the primary induces substantial fluctuations and gaps in the mass fallback rate, thereby manifesting as irregularities in the X-ray and UV/optical light curves. 

While the properties of AT2018fyk appear to be consistent with a SMBHB TDE, the inferred chirp mass is too low to be detected as a single PTA source. Conversely, we cannot dismiss the possibility of a repeating pTDE based on current observations alone. Additional observations are necessary to invalidate the SMBHB, the pTDE hypothesis, or both. The SMBHB model predicts a velocity change in the secondary BH of $\sim 5.5 \times 10^3 \rm{km~s^{-1}}$ over the $\approx$7.4 year binary orbital period. We attempted to detect regular red-shifts or blue-shifts in the emission lines using optical spectra. Unfortunately, only optical spectra taken during the peak of the initial flare are available, and the object has since become too dim to for spectroscopic ground based observations, making it challenging to verify this prediction currently. However, in principle, if there is a third flare or outburst, obtaining optical spectra might allow us to resolve any time-resolved offsets. 


\begin{acknowledgements}
We thank F. Liu and E.~R.~Coughlin for the useful discussion about binary TDEs. We thank B. Mockler for guidance in utilizing the MOSFiT package. This work made use of data supplied by the UK {\it Swift} Science Data Centre at the University of Leicester. This research has made use of data obtained through the High Energy Astrophysics Science Archive Research Center Online Service, provided by the NASA/Goddard Space Flight Center. Our work here is partly based on observations obtained with {\it XMM-Newton}, an ESA science mission with instruments and contributions directly funded by ESA Member States and NASA. SW
thanks the RU Department of Astrophysics/IMAPP for postdoctoral support during the early part of this work. SW is also supported by the NSFC (grant 12333004), the Strategic Priority Research Program of the Chinese Academy of Sciences (grant  XDB0550200), and the Strategic Pioneer Program on Space Science, Chinese Academy of Sciences (grant XDA15310300).
AIZ acknowledges 
support from NASA ADAP grant 80NSSC21K0988. PGJ has received funding from the European Research Council (ERC) under the
European Union’s Horizon 2020
research and innovation programme (Grant agreement No.~101095973).
\end{acknowledgements}

\begin{appendix}
\section{Constraints of $M_\bullet$ and \lowercase{ {$a_\bullet$}} using different X-ray spectra}
\label{app:1}

We fit the spectra {\it Swift}~1, {\it XMM}~1, {\it XMM}~2 and {\it XMM}~4, separately. For {\it Swift}~1 data, we fit the spectrum with a fit-function ${\tt TBAbs\times slimd}$, while for the three {\it XMM} spectra, we fit them with a fit-function  ${\tt TBAbs \times Thcomp(slimd)}$. For the model {\tt Thcomp}, we fix the electron temperature to a value of 300 keV. We derive the $M_\bullet$ and $a_\bullet$ contours for each spectrum with the command {\tt steppar} in {\sc XSPEC} \citep{Arnaud96}.

\begin{figure*}
    \centering
    \includegraphics[width=0.45\textwidth]{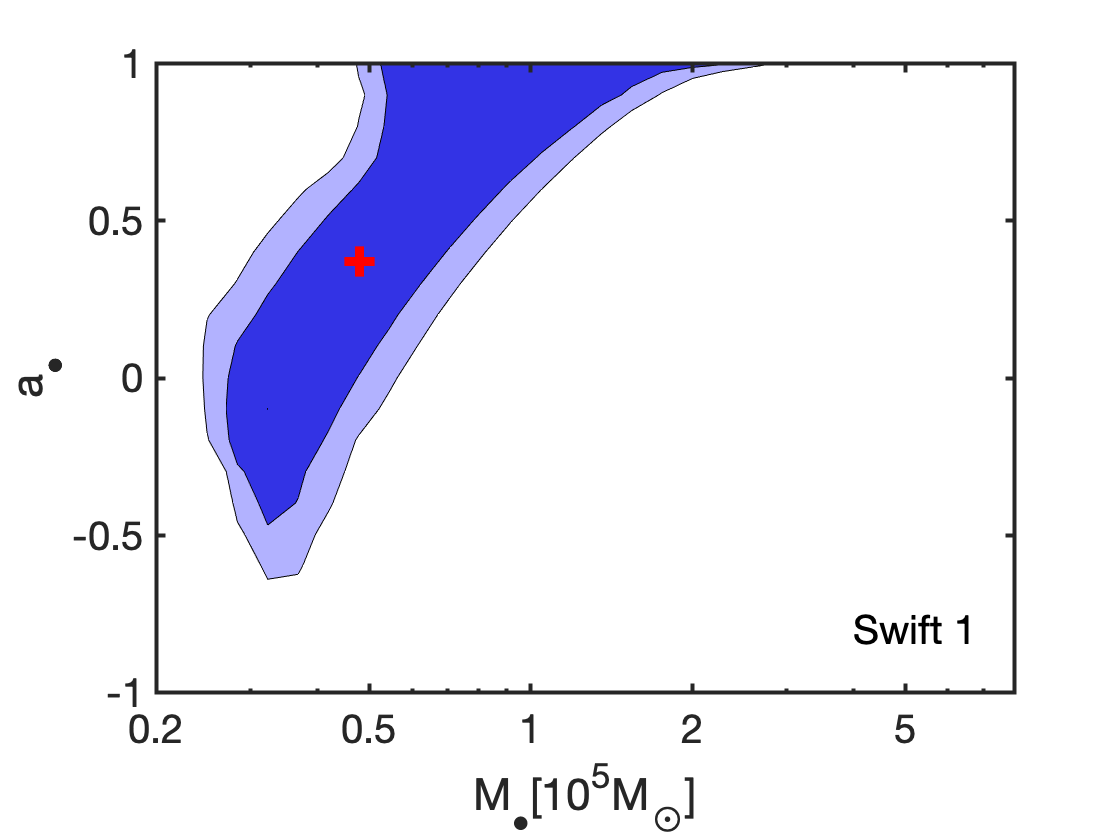}
    \includegraphics[width=0.45\textwidth]{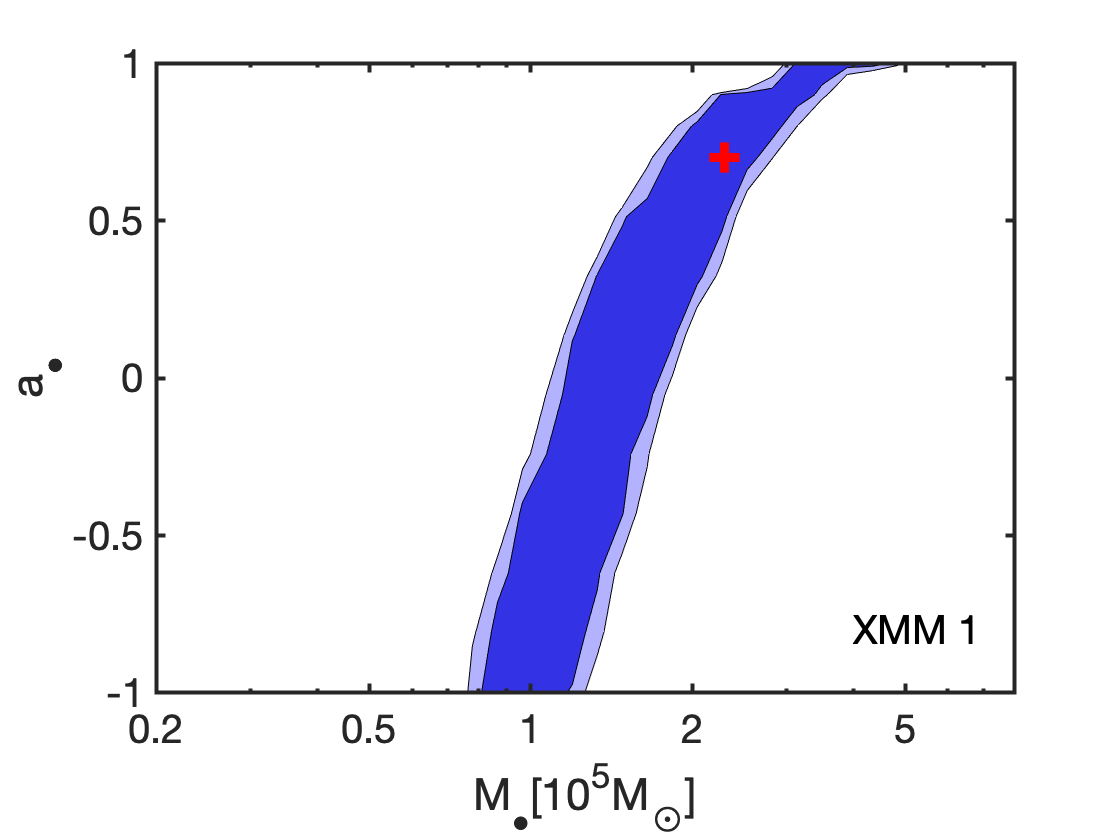}
    \includegraphics[width=0.45\textwidth]{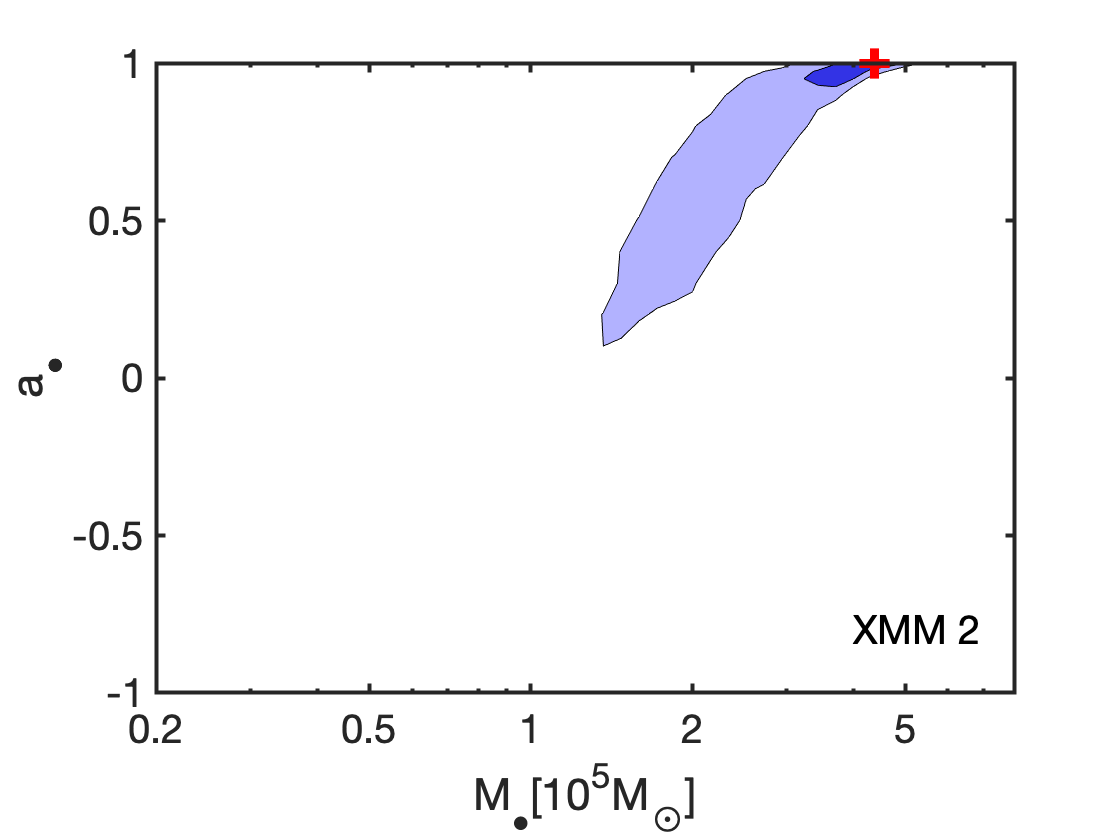}
    \includegraphics[width=0.45\textwidth]{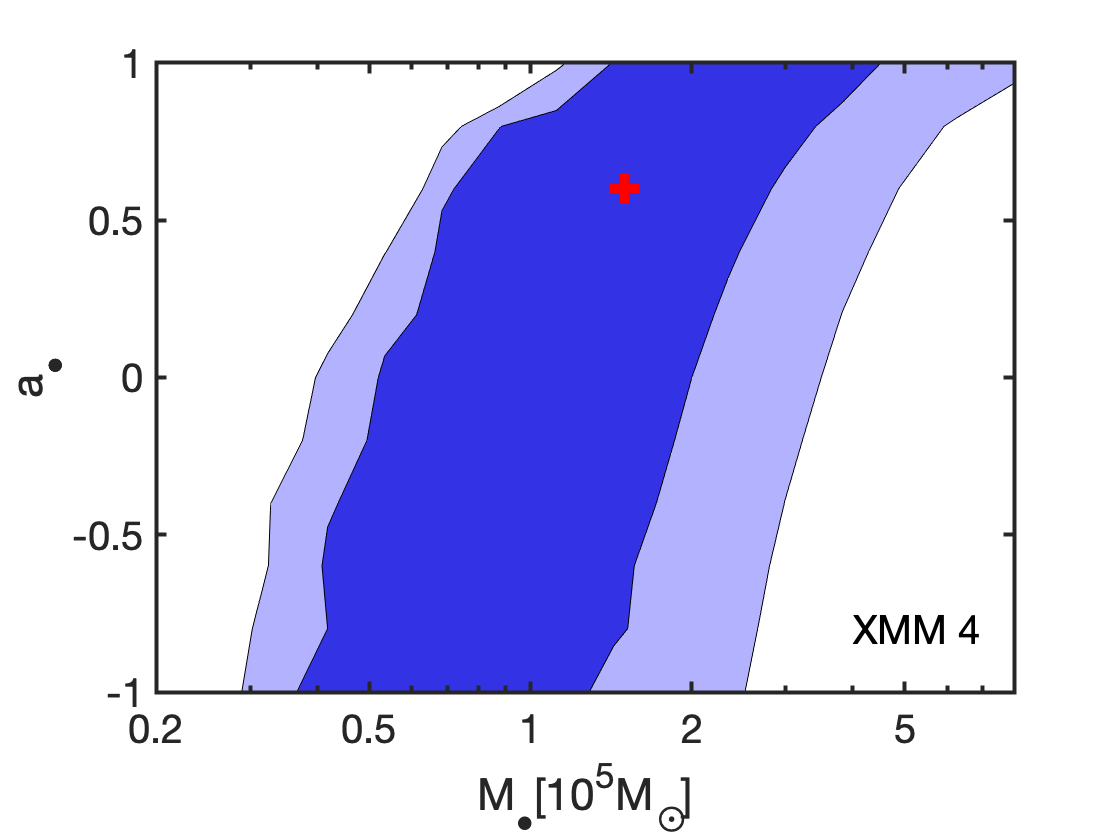}
    \caption{Constraints on $M_\bullet$ and $a_\bullet$ obtained by fitting different spectra individually. The dark and light blue regions denote the 1 and $2\sigma$ contours, respectively. {\it XMM}~2 yields a larger value for $M_\bullet$ than the three other epochs at the $>2\sigma$ level. }
    \label{fig:tension}
\end{figure*}

Figure~\ref{fig:tension} shows the fitting constraints on $M_\bullet$ and $a_\bullet$. The fits to the individual {\it XMM} spectra do not constrain $a_\bullet$ as much as the fit to the {\it Swift}~1 spectrum does, even though the individual {\it XMM} spectra have much more photons. This is because in the {\it XMM} epochs the corona modifies the spectrum, which is cloaking some information from the inner part of the disk. The constraints on $M_\bullet$ and $a_\bullet$ from {\it XMM}~2 are at least $2\sigma$ away from that of {\it Swift}~1. This tension may be explained by the possible soft lags in the {\it XMM}~2 as report by \cite{Zhang2022}. This tension becomes smaller for {\it XMM}~4. This may be because the outflowing material causing the soft lag disappears over time, or just because the there are not enough counts to give a robust constraints on $M_\bullet$ and $a_\bullet$ in the {\it XMM}~4 spectrum.

\section{Introduction of the {\tt slimd} model.}
\label{app:2}

 The {\tt slimd} model \citep{Wen_2022} is based on the work of \cite{Wen2020} and \cite{Wen_2021}, who used a general relativistic stationary slim disk model \citep{Abramowicz1996,Sadowski09} employing the ray-tracing code from \citet{JP11} to calculate the synthetic spectrum as seen by an observer.  The model assumptions include: 1. A circular and equatorial inner disk. 2. A zero-torque inner boundary condition, large optical depth, and absence of self-irradiation. 3. Generation of X-rays occur at the finite-height photospheric surface of the disk, presumed to match its scale height. The X-ray spectrum is determined by BH mass and spin, the mass accretion rate, the system inclination, and the spectral hardening factor $f_c$. The value of the latter parameter is determined through radiative transfer calculations by \cite{Davis2019}. Studies have indicated that different choices of the hardening factor have a limited impact on the estimation of mass and spin (see figures 14 and 15 in \citealt{Wen2020}). To study the constraining power on disk parameters of the thermal spectrum, we fit the {\tt Swift 1} spectrum with the {\tt slimd} model. 
\begin{figure}
    \centering
    \includegraphics[width=0.43\textwidth]{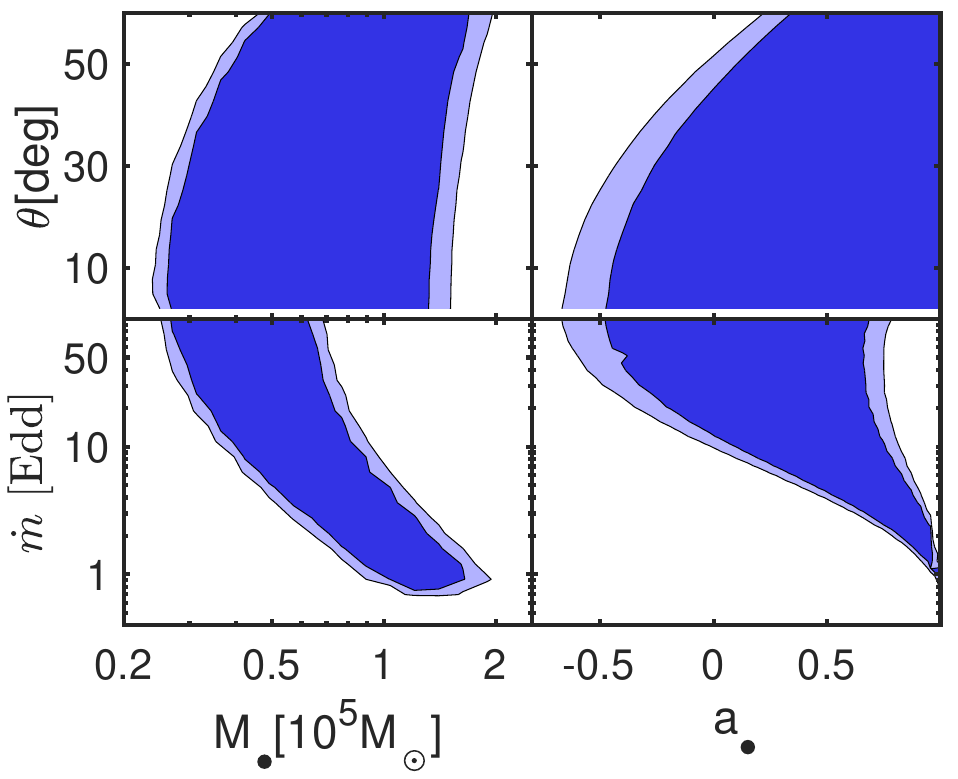}
    \caption{Constraints on different parameters obtained from fitting the {\tt Swift 1}. We fit the spectrum with the function ${\tt TBAbs\times slimd}$ by setting $N_{H}=1.15\times10^{20} \rm{cm^{-2}}$ while allowing the other parameters (including $M_\bullet$, $a_\bullet$, $\dot m$, and $\theta$) to float freely. The dark and light blue regions denote the 1 and $2\sigma$ contours, respectively.} 
    \label{fig:contour}
\end{figure}

Figure \ref{fig:contour} shows the degeneracy between disk parameters and the BH mass and spin. The BH mass and spin strongly (anti)correlate with the mass accretion rate and system inclination, respectively. There is no constraint on the inclination, and a poor constraint on the accretion rate. Although the uncertainty in the best-fit value of $\theta$ and $\dot m$ is large, the $M_\bullet$ and $a_\bullet$ values are relatively well constrained, as shown in the upper left panel of figure \ref{fig:tension}. Overall, the BH mass is better constrained than the BH spin. Due to the degeneracy between mass and spin, the constraint on spin is weak. 
However, the degeneracy between mass and spin can be partially alleviated by simultaneously fitting multi-epoch spectra with varying accretion rates \citep{Wen_2022}. These results indicate that the X-ray spectra are mainly determined by the BH mass and accretion rate and to a lesser degree BH spin. Spectra with super-Eddington accretion rates have a stronger power in constraining $M_\bullet$ and $a_\bullet$, because then the flux is insensitive to the actual value of the mass accretion rate \citep{Wen_2022}. 

\end{appendix}

\bibliography{bib}{}

\begin{thebibliography}{}
\expandafter\ifx\csname natexlab\endcsname\relax\def\natexlab#1{#1}\fi
\providecommand{\url}[1]{\href{#1}{#1}}
\providecommand{\dodoi}[1]{doi:~\href{http://doi.org/#1}{\nolinkurl{#1}}}
\providecommand{\doeprint}[1]{\href{http://ascl.net/#1}{\nolinkurl{http://ascl.net/#1}}}
\providecommand{\doarXiv}[1]{\href{https://arxiv.org/abs/#1}{\nolinkurl{https://arxiv.org/abs/#1}}}

\bibitem[{Abramowicz(2005)}]{Abramowicz2005}
Abramowicz, M. 2005, in Growing Black Holes: Accretion in a Cosmological
  Context, ed. A.~Merloni, S.~Nayakshin, \& R.~A. Sunyaev (Berlin, Heidelberg:
  Springer Berlin Heidelberg), 257--273

\bibitem[{{Abramowicz} {et~al.}(1996){Abramowicz}, {Chen}, {Granath}, \&
  {Lasota}}]{Abramowicz1996}
{Abramowicz}, M.~A., {Chen}, X.~M., {Granath}, M., \& {Lasota}, J.~P. 1996,
  \apj, 471, 762, \dodoi{10.1086/178004}

\bibitem[{{Abramowicz} {et~al.}(1988){Abramowicz}, {Czerny}, {Lasota}, \&
  {Szuszkiewicz}}]{Abramowicz1988}
{Abramowicz}, M.~A., {Czerny}, B., {Lasota}, J.~P., \& {Szuszkiewicz}, E. 1988,
  \apj, 332, 646, \dodoi{10.1086/166683}

\bibitem[{{Agazie} {et~al.}(2023){Agazie}, {Anumarlapudi}, {Archibald},
  {Arzoumanian}, {Baker}, {B{\'e}csy}, {Blecha}, {Brazier}, {Brook},
  {Burke-Spolaor}, {Burnette}, {Case}, {Charisi}, {Chatterjee},
  {Chatziioannou}, {Cheeseboro}, {Chen}, {Cohen}, {Cordes}, {Cornish},
  {Crawford}, {Cromartie}, {Crowter}, {Cutler}, {Decesar}, {Degan}, {Demorest},
  {Deng}, {Dolch}, {Drachler}, {Ellis}, {Ferrara}, {Fiore}, {Fonseca},
  {Freedman}, {Garver-Daniels}, {Gentile}, {Gersbach}, {Glaser}, {Good},
  {G{\"u}ltekin}, {Hazboun}, {Hourihane}, {Islo}, {Jennings}, {Johnson},
  {Jones}, {Kaiser}, {Kaplan}, {Kelley}, {Kerr}, {Key}, {Klein}, {Laal}, {Lam},
  {Lamb}, {Lazio}, {Lewandowska}, {Littenberg}, {Liu}, {Lommen}, {Lorimer},
  {Luo}, {Lynch}, {Ma}, {Madison}, {Mattson}, {McEwen}, {McKee}, {McLaughlin},
  {McMann}, {Meyers}, {Meyers}, {Mingarelli}, {Mitridate}, {Natarajan}, {Ng},
  {Nice}, {Ocker}, {Olum}, {Pennucci}, {Perera}, {Petrov}, {Pol}, {Radovan},
  {Ransom}, {Ray}, {Romano}, {Sardesai}, {Schmiedekamp}, {Schmiedekamp},
  {Schmitz}, {Schult}, {Shapiro-Albert}, {Siemens}, {Simon}, {Siwek}, {Stairs},
  {Stinebring}, {Stovall}, {Sun}, {Susobhanan}, {Swiggum}, {Taylor}, {Taylor},
  {Turner}, {Unal}, {Vallisneri}, {van Haasteren}, {Vigeland}, {Wahl}, {Wang},
  {Witt}, {Young}, \& {Nanograv Collaboration}}]{Agazie2023}
{Agazie}, G., {Anumarlapudi}, A., {Archibald}, A.~M., {et~al.} 2023, \apjl,
  951, L8, \dodoi{10.3847/2041-8213/acdac6}

\bibitem[{{Akaike}(1974)}]{Akaike1974}
{Akaike}, H. 1974, IEEE Transactions on Automatic Control, 19, 716

\bibitem[{{Amaro-Seoane} {et~al.}(2023){Amaro-Seoane}, {Andrews}, {Arca Sedda},
  {Askar}, {Baghi}, {Balasov}, {Bartos}, {Bavera}, {Bellovary}, {Berry},
  {Berti}, {Bianchi}, {Blecha}, {Blondin}, {Bogdanovi{\'c}}, {Boissier},
  {Bonetti}, {Bonoli}, {Bortolas}, {Breivik}, {Capelo}, {Caramete},
  {Cattorini}, {Charisi}, {Chaty}, {Chen}, {Chru{\'s}li{\'n}ska}, {Chua},
  {Church}, {Colpi}, {D'Orazio}, {Danielski}, {Davies}, {Dayal}, {De Rosa},
  {Derdzinski}, {Destounis}, {Dotti}, {Dutan}, {Dvorkin}, {Fabj}, {Foglizzo},
  {Ford}, {Fouvry}, {Franchini}, {Fragos}, {Fryer}, {Gaspari}, {Gerosa},
  {Graziani}, {Groot}, {Habouzit}, {Haggard}, {Haiman}, {Han}, {Istrate},
  {Johansson}, {Khan}, {Kimpson}, {Kokkotas}, {Kong}, {Korol}, {Kremer},
  {Kupfer}, {Lamberts}, {Larson}, {Lau}, {Liu}, {Lloyd-Ronning}, {Lodato},
  {Lupi}, {Ma}, {Maccarone}, {Mandel}, {Mangiagli}, {Mapelli}, {Mathis},
  {Mayer}, {McGee}, {McKernan}, {Miller}, {Mota}, {Mumpower}, {Nasim},
  {Nelemans}, {Noble}, {Pacucci}, {Panessa}, {Paschalidis}, {Pfister},
  {Porquet}, {Quenby}, {Ricarte}, {R{\"o}pke}, {Regan}, {Rosswog}, {Ruiter},
  {Ruiz}, {Runnoe}, {Schneider}, {Schnittman}, {Secunda}, {Sesana}, {Seto},
  {Shao}, {Shapiro}, {Sopuerta}, {Stone}, {Suvorov}, {Tamanini}, {Tamfal},
  {Tauris}, {Temmink}, {Tomsick}, {Toonen}, {Torres-Orjuela}, {Toscani},
  {Tsokaros}, {Unal}, {V{\'a}zquez-Aceves}, {Valiante}, {van Putten}, {van
  Roestel}, {Vignali}, {Volonteri}, {Wu}, {Younsi}, {Yu}, {Zane}, {Zwick},
  {Antonini}, {Baibhav}, {Barausse}, {Bonilla Rivera}, {Branchesi},
  {Branduardi-Raymont}, {Burdge}, {Chakraborty}, {Cuadra}, {Dage}, {Davis}, {de
  Mink}, {Decarli}, {Doneva}, {Escoffier}, {Gandhi}, {Haardt}, {Lousto},
  {Nissanke}, {Nordhaus}, {O'Shaughnessy}, {Portegies Zwart}, {Pound},
  {Schussler}, {Sergijenko}, {Spallicci}, {Vernieri}, \&
  {Vigna-G{\'o}mez}}]{LISA2023}
{Amaro-Seoane}, P., {Andrews}, J., {Arca Sedda}, M., {et~al.} 2023, Living
  Reviews in Relativity, 26, 2, \dodoi{10.1007/s41114-022-00041-y}

\bibitem[{{Andalman} {et~al.}(2022){Andalman}, {Liska}, {Tchekhovskoy},
  {Coughlin}, \& {Stone}}]{Andalman+22}
{Andalman}, Z.~L., {Liska}, M. T.~P., {Tchekhovskoy}, A., {Coughlin}, E.~R., \&
  {Stone}, N. 2022, \mnras, 510, 1627, \dodoi{10.1093/mnras/stab3444}

\bibitem[{{Antoniadis} {et~al.}(2023){Antoniadis}, {Arumugam}, {Arumugam},
  {Babak}, {Bagchi}, {Bak Nielsen}, {Bassa}, {Bathula}, {Berthereau},
  {Bonetti}, {Bortolas}, {Brook}, {Burgay}, {Caballero}, {Chalumeau},
  {Champion}, {Chanlaridis}, {Chen}, {Cognard}, {Dandapat}, {Deb}, {Desai},
  {Desvignes}, {Dhanda-Batra}, {Dwivedi}, {Falxa}, {Ferdman}, {Franchini},
  {Gair}, {Goncharov}, {Gopakumar}, {Graikou}, {Grie{\ss}meier}, {Guillemot},
  {Guo}, {Gupta}, {Hisano}, {Hu}, {Iraci}, {Izquierdo-Villalba}, {Jang},
  {Jawor}, {Janssen}, {Jessner}, {Joshi}, {Kareem}, {Karuppusamy}, {Keane},
  {Keith}, {Kharbanda}, {Kikunaga}, {Kolhe}, {Kramer}, {Krishnakumar},
  {Lackeos}, {Lee}, {Liu}, {Liu}, {Lyne}, {McKee}, {Maan}, {Main},
  {Mickaliger}, {Nitu}, {Nobleson}, {Paladi}, {Parthasarathy}, {Perera},
  {Perrodin}, {Petiteau}, {Porayko}, {Possenti}, {Prabu}, {Quelquejay Leclere},
  {Rana}, {Samajdar}, {Sanidas}, {Sesana}, {Shaifullah}, {Singha}, {Speri},
  {Spiewak}, {Srivastava}, {Stappers}, {Surnis}, {Susarla}, {Susobhanan},
  {Takahashi}, {Tarafdar}, {Theureau}, {Tiburzi}, {van der Wateren}, {Vecchio},
  {Venkatraman Krishnan}, {Verbiest}, {Wang}, {Wang}, \& {Wu}}]{Antoniadis2023}
{Antoniadis}, J., {Arumugam}, P., {Arumugam}, S., {et~al.} 2023, arXiv
  e-prints, arXiv:2306.16214, \dodoi{10.48550/arXiv.2306.16214}

\bibitem[{{Arcavi} {et~al.}(2014){Arcavi}, {Gal-Yam}, {Sullivan}, {Pan},
  {Cenko}, {Horesh}, {Ofek}, {De Cia}, {Yan}, {Yang}, {Howell}, {Tal},
  {Kulkarni}, {Tendulkar}, {Tang}, {Xu}, {Sternberg}, {Cohen}, {Bloom},
  {Nugent}, {Kasliwal}, {Perley}, {Quimby}, {Miller}, {Theissen}, \&
  {Laher}}]{Arcavi2014}
{Arcavi}, I., {Gal-Yam}, A., {Sullivan}, M., {et~al.} 2014, \apj, 793, 38,
  \dodoi{10.1088/0004-637X/793/1/38}

\bibitem[{{Arnaud}(1996)}]{Arnaud96}
{Arnaud}, K.~A. 1996, in Astronomical Society of the Pacific Conference Series,
  Vol. 101, Astronomical Data Analysis Software and Systems V, ed. G.~H.
  {Jacoby} \& J.~{Barnes}, 17

\bibitem[{{Begelman} {et~al.}(1980){Begelman}, {Blandford}, \&
  {Rees}}]{Begelman1980}
{Begelman}, M.~C., {Blandford}, R.~D., \& {Rees}, M.~J. 1980, \nat, 287, 307,
  \dodoi{10.1038/287307a0}

\bibitem[{{Blandford} \& {Rees}(1974)}]{Blandford1974}
{Blandford}, R.~D., \& {Rees}, M.~J. 1974, \mnras, 169, 395,
  \dodoi{10.1093/mnras/169.3.395}

\bibitem[{{Bonnerot} \& {Lu}(2020)}]{BonnerotLu19}
{Bonnerot}, C., \& {Lu}, W. 2020, \mnras, 495, 1374,
  \dodoi{10.1093/mnras/staa1246}

\bibitem[{{Cannizzo} {et~al.}(1990){Cannizzo}, {Lee}, \&
  {Goodman}}]{Cannizzo+90}
{Cannizzo}, J.~K., {Lee}, H.~M., \& {Goodman}, J. 1990, \apj, 351, 38,
  \dodoi{10.1086/168442}

\bibitem[{{Cao} {et~al.}(2023){Cao}, {Jonker}, {Wen}, {Stone}, \&
  {Zabludoff}}]{Cao2023}
{Cao}, Z., {Jonker}, P.~G., {Wen}, S., {Stone}, N.~C., \& {Zabludoff}, A.~I.
  2023, \mnras, 519, 2375, \dodoi{10.1093/mnras/stac3539}

\bibitem[{{Cash}(1979)}]{Cash79}
{Cash}, W. 1979, \apj, 228, 939, \dodoi{10.1086/156922}

\bibitem[{{Chen} {et~al.}(2011){Chen}, {Sesana}, {Madau}, \& {Liu}}]{Chen2011}
{Chen}, X., {Sesana}, A., {Madau}, P., \& {Liu}, F.~K. 2011, \apj, 729, 13,
  \dodoi{10.1088/0004-637X/729/1/13}

\bibitem[{{Chiang} {et~al.}(2015){Chiang}, {Walton}, {Fabian}, {Wilkins}, \&
  {Gallo}}]{Chiang15}
{Chiang}, C.-Y., {Walton}, D.~J., {Fabian}, A.~C., {Wilkins}, D.~R., \&
  {Gallo}, L.~C. 2015, \mnras, 446, 759, \dodoi{10.1093/mnras/stu2087}

\bibitem[{{Cohn} \& {Kulsrud}(1978)}]{CohnKulsrud78}
{Cohn}, H., \& {Kulsrud}, R.~M. 1978, \apj, 226, 1087, \dodoi{10.1086/156685}

\bibitem[{{Coughlin} \& {Armitage}(2018)}]{Coughlin2018}
{Coughlin}, E.~R., \& {Armitage}, P.~J. 2018, \mnras, 474, 3857,
  \dodoi{10.1093/mnras/stx3039}

\bibitem[{{Coughlin} {et~al.}(2019){Coughlin}, {Armitage}, {Lodato}, \&
  {Nixon}}]{Coughlin2019}
{Coughlin}, E.~R., {Armitage}, P.~J., {Lodato}, G., \& {Nixon}, C.~J. 2019,
  \ssr, 215, 45, \dodoi{10.1007/s11214-019-0612-z}

\bibitem[{{Coughlin} {et~al.}(2017){Coughlin}, {Armitage}, {Nixon}, \&
  {Begelman}}]{Coughlin2017}
{Coughlin}, E.~R., {Armitage}, P.~J., {Nixon}, C., \& {Begelman}, M.~C. 2017,
  \mnras, 465, 3840, \dodoi{10.1093/mnras/stw2913}

\bibitem[{{Coughlin} \& {Nixon}(2019)}]{Coughlin2019b}
{Coughlin}, E.~R., \& {Nixon}, C.~J. 2019, \apjl, 883, L17,
  \dodoi{10.3847/2041-8213/ab412d}

\bibitem[{{Coughlin} \& {Nixon}(2022)}]{Coughlin2022}
---. 2022, \apj, 936, 70, \dodoi{10.3847/1538-4357/ac85b3}

\bibitem[{{Dai} {et~al.}(2015){Dai}, {McKinney}, \& {Miller}}]{Dai15}
{Dai}, L., {McKinney}, J.~C., \& {Miller}, M.~C. 2015, \apjl, 812, L39,
  \dodoi{10.1088/2041-8205/812/2/L39}

\bibitem[{{Dai} {et~al.}(2018){Dai}, {McKinney}, {Roth}, {Ramirez-Ruiz}, \&
  {Miller}}]{Dai2018}
{Dai}, L., {McKinney}, J.~C., {Roth}, N., {Ramirez-Ruiz}, E., \& {Miller},
  M.~C. 2018, \apjl, 859, L20, \dodoi{10.3847/2041-8213/aab429}

\bibitem[{{Davis} \& {El-Abd}(2019)}]{Davis2019}
{Davis}, S.~W., \& {El-Abd}, S. 2019, \apj, 874, 23,
  \dodoi{10.3847/1538-4357/ab05c5}

\bibitem[{{De Rosa} {et~al.}(2019){De Rosa}, {Vignali}, {Bogdanovi{\'c}},
  {Capelo}, {Charisi}, {Dotti}, {Husemann}, {Lusso}, {Mayer}, {Paragi},
  {Runnoe}, {Sesana}, {Steinborn}, {Bianchi}, {Colpi}, {del Valle}, {Frey},
  {Gab{\'a}nyi}, {Giustini}, {Guainazzi}, {Haiman}, {Herrera Ruiz},
  {Herrero-Illana}, {Iwasawa}, {Komossa}, {Lena}, {Loiseau}, {Perez-Torres},
  {Piconcelli}, \& {Volonteri}}]{DeRosa2019}
{De Rosa}, A., {Vignali}, C., {Bogdanovi{\'c}}, T., {et~al.} 2019, \nar, 86,
  101525, \dodoi{10.1016/j.newar.2020.101525}

\bibitem[{{Donato} {et~al.}(2014){Donato}, {Cenko}, {Covino}, {Troja},
  {Pursimo}, {Cheung}, {Fox}, {Kutyrev}, {Campana}, {Fugazza}, {Landt}, \&
  {Butler}}]{Donato2014}
{Donato}, D., {Cenko}, S.~B., {Covino}, S., {et~al.} 2014, \apj, 781, 59,
  \dodoi{10.1088/0004-637X/781/2/59}

\bibitem[{{Done} \& {Gierli{\'n}ski}(2005)}]{Done2005}
{Done}, C., \& {Gierli{\'n}ski}, M. 2005, \mnras, 364, 208,
  \dodoi{10.1111/j.1365-2966.2005.09555.x}

\bibitem[{Evans {et~al.}(2009)Evans, Beardmore, Page, Osborne, O'Brien,
  Willingale, Starling, Burrows, Godet, Vetere, Racusin, Goad, Wiersema,
  Angelini, Capalbi, Chincarini, Gehrels, Kennea, Margutti, Morris, Mountford,
  Pagani, Perri, Romano, \& Tanvir}]{Evans2009}
Evans, P.~A., Beardmore, A.~P., Page, K.~L., {et~al.} 2009, Monthly Notices of
  the Royal Astronomical Society, 397, 1177,
  \dodoi{10.1111/j.1365-2966.2009.14913.x}

\bibitem[{{Ferrarese} \& {Ford}(2005)}]{Ferrarese2005}
{Ferrarese}, L., \& {Ford}, H. 2005, \ssr, 116, 523,
  \dodoi{10.1007/s11214-005-3947-6}

\bibitem[{{Ford} {et~al.}(2000){Ford}, {Kozinsky}, \& {Rasio}}]{Ford+00}
{Ford}, E.~B., {Kozinsky}, B., \& {Rasio}, F.~A. 2000, \apj, 535, 385,
  \dodoi{10.1086/308815}

\bibitem[{{Frank} \& {Rees}(1976)}]{FrankRees76}
{Frank}, J., \& {Rees}, M.~J. 1976, \mnras, 176, 633,
  \dodoi{10.1093/mnras/176.3.633}

\bibitem[{{French} {et~al.}(2016){French}, {Arcavi}, \&
  {Zabludoff}}]{French2016}
{French}, K.~D., {Arcavi}, I., \& {Zabludoff}, A. 2016, \apjl, 818, L21,
  \dodoi{10.3847/2041-8205/818/1/L21}

\bibitem[{{Gaskell}(1983)}]{Gaskell1983}
{Gaskell}, C.~M. 1983, in Liege International Astrophysical Colloquia, Vol.~24,
  Liege International Astrophysical Colloquia, ed. J.-P. {Swings}, 473--477

\bibitem[{{Gelman} \& {Rubin}(1992)}]{Gelman1992}
{Gelman}, A., \& {Rubin}, D.~B. 1992, Statistical Science, 7, 457,
  \dodoi{10.1214/ss/1177011136}

\bibitem[{{Goldtooth} {et~al.}(2023){Goldtooth}, {Zabludoff}, {Wen}, {Jonker},
  {Stone}, \& {Cao}}]{Goldtooth2023}
{Goldtooth}, A., {Zabludoff}, A.~I., {Wen}, S., {et~al.} 2023, \pasp, 135,
  034101, \dodoi{10.1088/1538-3873/acb9bc}

\bibitem[{{Gong} {et~al.}(2021){Gong}, {Luo}, \& {Wang}}]{Tianqin2021}
{Gong}, Y., {Luo}, J., \& {Wang}, B. 2021, Nature Astronomy, 5, 881,
  \dodoi{10.1038/s41550-021-01480-3}

\bibitem[{{Gordon} {et~al.}(2003){Gordon}, {Clayton}, {Misselt}, {Landolt}, \&
  {Wolff}}]{Gordon2003}
{Gordon}, K.~D., {Clayton}, G.~C., {Misselt}, K.~A., {Landolt}, A.~U., \&
  {Wolff}, M.~J. 2003, \apj, 594, 279, \dodoi{10.1086/376774}

\bibitem[{{Graur} {et~al.}(2018){Graur}, {French}, {Zahid}, {Guillochon},
  {Mandel}, {Auchettl}, \& {Zabludoff}}]{Graur2018}
{Graur}, O., {French}, K.~D., {Zahid}, H.~J., {et~al.} 2018, \apj, 853, 39,
  \dodoi{10.3847/1538-4357/aaa3fd}

\bibitem[{{Greene} {et~al.}(2016){Greene}, {Seth}, {Kim}, {L{\"a}sker},
  {Goulding}, {Gao}, {Braatz}, {Henkel}, {Condon}, {Lo}, \&
  {Zhao}}]{Greene2016}
{Greene}, J.~E., {Seth}, A., {Kim}, M., {et~al.} 2016, \apjl, 826, L32,
  \dodoi{10.3847/2041-8205/826/2/L32}

\bibitem[{Guillochon {et~al.}(2014)Guillochon, Manukian, \&
  Ramirez-Ruiz}]{Guillochon+14}
Guillochon, J., Manukian, H., \& Ramirez-Ruiz, E. 2014, \apj, 783, 23,
  \dodoi{10.1088/0004-637x/783/1/23}

\bibitem[{{Guillochon} \& {Ramirez-Ruiz}(2013)}]{Guillochon2013}
{Guillochon}, J., \& {Ramirez-Ruiz}, E. 2013, \apj, 767, 25,
  \dodoi{10.1088/0004-637X/767/1/25}

\bibitem[{{Guo} {et~al.}(2023){Guo}, {Sun}, {Li}, {Jiang}, {Wang}, {Bu},
  {Jiang}, {Wang}, {Yao}, {Shen}, {Gu}, \& {Sun}}]{Guo2023}
{Guo}, H., {Sun}, J., {Li}, S.-L., {et~al.} 2023, arXiv e-prints,
  arXiv:2312.06771, \dodoi{10.48550/arXiv.2312.06771}

\bibitem[{{Hammerstein} {et~al.}(2023){Hammerstein}, {van Velzen}, {Gezari},
  {Cenko}, {Yao}, {Ward}, {Frederick}, {Villanueva}, {Somalwar}, {Graham},
  {Kulkarni}, {Stern}, {Andreoni}, {Bellm}, {Dekany}, {Dhawan}, {Drake},
  {Fremling}, {Gatkine}, {Groom}, {Ho}, {Kasliwal}, {Karambelkar}, {Kool},
  {Masci}, {Medford}, {Perley}, {Purdum}, {van Roestel}, {Sharma}, {Sollerman},
  {Taggart}, \& {Yan}}]{Hammerstein2022}
{Hammerstein}, E., {van Velzen}, S., {Gezari}, S., {et~al.} 2023, \apj, 942, 9,
  \dodoi{10.3847/1538-4357/aca283}

\bibitem[{{H{\"a}ring} \& {Rix}(2004)}]{Haring2004}
{H{\"a}ring}, N., \& {Rix}, H.-W. 2004, \apjl, 604, L89, \dodoi{10.1086/383567}

\bibitem[{{Hills}(1975)}]{Hills1975}
{Hills}, J.~G. 1975, \nat, 254, 295, \dodoi{10.1038/254295a0}

\bibitem[{{Ivanov} {et~al.}(2005){Ivanov}, {Polnarev}, \& {Saha}}]{Ivanov2005}
{Ivanov}, P.~B., {Polnarev}, A.~G., \& {Saha}, P. 2005, \mnras, 358, 1361,
  \dodoi{10.1111/j.1365-2966.2005.08843.x}

\bibitem[{{Jansen} {et~al.}(2001){Jansen}, {Lumb}, {Altieri}, {Clavel}, {Ehle},
  {Erd}, {Gabriel}, {Guainazzi}, {Gondoin}, {Much}, {Munoz}, {Santos},
  {Schartel}, {Texier}, \& {Vacanti}}]{Jansen2001}
{Jansen}, F., {Lumb}, D., {Altieri}, B., {et~al.} 2001, \aap, 365, L1,
  \dodoi{10.1051/0004-6361:20000036}

\bibitem[{{Kormendy} \& {Ho}(2013)}]{Kormendy2013}
{Kormendy}, J., \& {Ho}, L.~C. 2013, \araa, 51, 511,
  \dodoi{10.1146/annurev-astro-082708-101811}

\bibitem[{{Koss} {et~al.}(2022){Koss}, {Trakhtenbrot}, {Ricci}, {Oh}, {Bauer},
  {Stern}, {Caglar}, {den Brok}, {Mushotzky}, {Ricci}, {Mej{\'\i}a-Restrepo},
  {Lamperti}, {Treister}, {B{\"a}r}, {Harrison}, {Powell}, {Privon}, {Riffel},
  {Rojas}, {Schawinski}, \& {Urry}}]{Koss2022}
{Koss}, M.~J., {Trakhtenbrot}, B., {Ricci}, C., {et~al.} 2022, \apjs, 261, 6,
  \dodoi{10.3847/1538-4365/ac650b}

\bibitem[{{Kozai}(1962)}]{Kozai1962}
{Kozai}, Y. 1962, \aj, 67, 591, \dodoi{10.1086/108790}

\bibitem[{{Lidov}(1962)}]{Lidov1962}
{Lidov}, M.~L. 1962, \planss, 9, 719, \dodoi{10.1016/0032-0633(62)90129-0}

\bibitem[{{Lightman} \& {Shapiro}(1977)}]{LightmanShapiro77}
{Lightman}, A.~P., \& {Shapiro}, S.~L. 1977, \apj, 211, 244,
  \dodoi{10.1086/154925}

\bibitem[{{Lin} {et~al.}(2017){Lin}, {Guillochon}, {Komossa}, {Ramirez-Ruiz},
  {Irwin}, {Maksym}, {Grupe}, {Godet}, {Webb}, {Barret}, {Zauderer}, {Duc},
  {Carrasco}, \& {Gwyn}}]{Lin2017}
{Lin}, D., {Guillochon}, J., {Komossa}, S., {et~al.} 2017, Nature Astronomy, 1,
  0033, \dodoi{10.1038/s41550-016-0033}

\bibitem[{{Lin} {et~al.}(2018){Lin}, {Strader}, {Carrasco}, {Page},
  {Romanowsky}, {Homan}, {Irwin}, {Remillard}, {Godet}, {Webb}, {Baumgardt},
  {Wijnands}, {Barret}, {Duc}, {Brodie}, \& {Gwyn}}]{Lin2018}
{Lin}, D., {Strader}, J., {Carrasco}, E.~R., {et~al.} 2018, Nature Astronomy,
  2, 656, \dodoi{10.1038/s41550-018-0493-1}

\bibitem[{{Liska} {et~al.}(2022){Liska}, {Musoke}, {Tchekhovskoy}, {Porth}, \&
  {Beloborodov}}]{Liska+22}
{Liska}, M.~T.~P., {Musoke}, G., {Tchekhovskoy}, A., {Porth}, O., \&
  {Beloborodov}, A.~M. 2022, \apjl, 935, L1, \dodoi{10.3847/2041-8213/ac84db}

\bibitem[{{Liu} {et~al.}(2021){Liu}, {Cao}, {Abramowicz}, {Wielgus}, {Cao}, \&
  {Zhou}}]{Liu_2021}
{Liu}, F.~K., {Cao}, C.~Y., {Abramowicz}, M.~A., {et~al.} 2021, \apj, 908, 179,
  \dodoi{10.3847/1538-4357/abd2b6}

\bibitem[{{Liu} {et~al.}(2009){Liu}, {Li}, \& {Chen}}]{Liu_2009}
{Liu}, F.~K., {Li}, S., \& {Chen}, X. 2009, \apjl, 706, L133,
  \dodoi{10.1088/0004-637X/706/1/L133}

\bibitem[{{Liu} {et~al.}(2014){Liu}, {Li}, \& {Komossa}}]{Liu2014}
{Liu}, F.~K., {Li}, S., \& {Komossa}, S. 2014, \apj, 786, 103,
  \dodoi{10.1088/0004-637X/786/2/103}

\bibitem[{{Liu} {et~al.}(2022){Liu}, {Dou}, {Chen}, \& {Shen}}]{Liu2022ApJ}
{Liu}, X.-L., {Dou}, L.-M., {Chen}, J.-H., \& {Shen}, R.-F. 2022, \apj, 925,
  67, \dodoi{10.3847/1538-4357/ac33a9}

\bibitem[{{Lodato} {et~al.}(2009){Lodato}, {Nayakshin}, {King}, \&
  {Pringle}}]{Lodato2009}
{Lodato}, G., {Nayakshin}, S., {King}, A.~R., \& {Pringle}, J.~E. 2009, \mnras,
  398, 1392, \dodoi{10.1111/j.1365-2966.2009.15179.x}

\bibitem[{{Loeb} \& {Ulmer}(1997)}]{Loeb+97}
{Loeb}, A., \& {Ulmer}, A. 1997, \apj, 489, 573, \dodoi{10.1086/304814}

\bibitem[{Lu \& Bonnerot(2019)}]{LuBonnerot19}
Lu, W., \& Bonnerot, C. 2019, \mnras, 492, 686, \dodoi{10.1093/mnras/stz3405}

\bibitem[{{Magorrian} \& {Tremaine}(1999)}]{Magorrian1999}
{Magorrian}, J., \& {Tremaine}, S. 1999, \mnras, 309, 447,
  \dodoi{10.1046/j.1365-8711.1999.02853.x}

\bibitem[{{Mayer} {et~al.}(2007){Mayer}, {Kazantzidis}, {Madau}, {Colpi},
  {Quinn}, \& {Wadsley}}]{Mayer2007}
{Mayer}, L., {Kazantzidis}, S., {Madau}, P., {et~al.} 2007, Science, 316, 1874,
  \dodoi{10.1126/science.1141858}

\bibitem[{{McConnell} \& {Ma}(2013)}]{Mcconnell2013}
{McConnell}, N.~J., \& {Ma}, C.-P. 2013, \apj, 764, 184,
  \dodoi{10.1088/0004-637X/764/2/184}

\bibitem[{{Merritt} \& {Milosavljevi{\'c}}(2005)}]{Merritt2005}
{Merritt}, D., \& {Milosavljevi{\'c}}, M. 2005, Living Reviews in Relativity,
  8, 8, \dodoi{10.12942/lrr-2005-8}

\bibitem[{{Merritt} \& {Poon}(2004)}]{Merritt2004}
{Merritt}, D., \& {Poon}, M.~Y. 2004, \apj, 606, 788, \dodoi{10.1086/382497}

\bibitem[{{Metzger}(2022)}]{Metzger2022}
{Metzger}, B.~D. 2022, \apjl, 937, L12, \dodoi{10.3847/2041-8213/ac90ba}

\bibitem[{Metzger \& Stone(2016)}]{Metzger2016}
Metzger, B.~D., \& Stone, N.~C. 2016, \mnras, 461, 948,
  \dodoi{10.1093/mnras/stw1394}

\bibitem[{{Miller} {et~al.}(2015){Miller}, {Kaastra}, {Miller}, {Reynolds},
  {Brown}, {Cenko}, {Drake}, {Gezari}, {Guillochon}, {Gultekin}, {Irwin},
  {Levan}, {Maitra}, {Maksym}, {Mushotzky}, {O'Brien}, {Paerels}, {de Plaa},
  {Ramirez-Ruiz}, {Strohmayer}, \& {Tanvir}}]{Miller2015}
{Miller}, J.~M., {Kaastra}, J.~S., {Miller}, M.~C., {et~al.} 2015, \nat, 526,
  542, \dodoi{10.1038/nature15708}

\bibitem[{{Milosavljevi{\'c}} \& {Merritt}(2003)}]{Milosavljevi2003}
{Milosavljevi{\'c}}, M., \& {Merritt}, D. 2003, \apj, 596, 860,
  \dodoi{10.1086/378086}

\bibitem[{{Mockler} {et~al.}(2019){Mockler}, {Guillochon}, \&
  {Ramirez-Ruiz}}]{Mockler19}
{Mockler}, B., {Guillochon}, J., \& {Ramirez-Ruiz}, E. 2019, \apj, 872, 151,
  \dodoi{10.3847/1538-4357/ab010f}

\bibitem[{{Mockler} {et~al.}(2023){Mockler}, {Melchor}, {Naoz}, \&
  {Ramirez-Ruiz}}]{Mockler2023}
{Mockler}, B., {Melchor}, D., {Naoz}, S., \& {Ramirez-Ruiz}, E. 2023, arXiv
  e-prints, arXiv:2306.05510, \dodoi{10.48550/arXiv.2306.05510}

\bibitem[{Mummery(2021)}]{Mummery2021tde}
Mummery, A. 2021, Monthly Notices of the Royal Astronomical Society: Letters,
  507, L24, \dodoi{10.1093/mnrasl/slab088}

\bibitem[{{Mummery} {et~al.}(2024){Mummery}, {van Velzen}, {Nathan}, {Ingram},
  {Hammerstein}, {Fraser-Taliente}, \& {Balbus}}]{Mummery2023b}
{Mummery}, A., {van Velzen}, S., {Nathan}, E., {et~al.} 2024, \mnras, 527,
  2452, \dodoi{10.1093/mnras/stad3001}

\bibitem[{{Mummery} {et~al.}(2023){Mummery}, {Wevers}, {Saxton}, \&
  {Pasham}}]{Mummery+23}
{Mummery}, A., {Wevers}, T., {Saxton}, R., \& {Pasham}, D. 2023, \mnras, 519,
  5828, \dodoi{10.1093/mnras/stac3798}

\bibitem[{{Naoz}(2016)}]{Naoz2016}
{Naoz}, S. 2016, \araa, 54, 441, \dodoi{10.1146/annurev-astro-081915-023315}

\bibitem[{{Pasham} {et~al.}(2020){Pasham}, {Gendreau}, {Remillard},
  {Loewenstein}, {Arzoumanian}, \& {Kara}}]{Pasham2020}
{Pasham}, D., {Gendreau}, K., {Remillard}, R., {et~al.} 2020, The Astronomer's
  Telegram, 13864, 1

\bibitem[{{Pasham} {et~al.}(2021){Pasham}, {Ho}, {Alston}, {Remillard}, {Ng},
  {Gendreau}, {Metzger}, {Altamirano}, {Chakrabarty}, {Fabian}, {Miller},
  {Bult}, {Arzoumanian}, {Steiner}, {Strohmayer}, {Tombesi}, {Homan},
  {Cackett}, \& {Harding}}]{pashcow}
{Pasham}, D.~R., {Ho}, W. C.~G., {Alston}, W., {et~al.} 2021, Nature Astronomy,
  6, 249, \dodoi{10.1038/s41550-021-01524-8}

\bibitem[{{Peters}(1964)}]{Peters1964}
{Peters}, P.~C. 1964, Physical Review, 136, 1224,
  \dodoi{10.1103/PhysRev.136.B1224}

\bibitem[{{Peterson} {et~al.}(1998){Peterson}, {Wanders}, {Horne}, {Collier},
  {Alexander}, {Kaspi}, \& {Maoz}}]{Peterson_1998}
{Peterson}, B.~M., {Wanders}, I., {Horne}, K., {et~al.} 1998, \pasp, 110, 660,
  \dodoi{10.1086/316177}

\bibitem[{{Peterson} {et~al.}(2004){Peterson}, {Ferrarese}, {Gilbert}, {Kaspi},
  {Malkan}, {Maoz}, {Merritt}, {Netzer}, {Onken}, {Pogge}, {Vestergaard}, \&
  {Wandel}}]{Peterson_2004}
{Peterson}, B.~M., {Ferrarese}, L., {Gilbert}, K.~M., {et~al.} 2004, \apj, 613,
  682, \dodoi{10.1086/423269}

\bibitem[{{Pfister} {et~al.}(2020){Pfister}, {Volonteri}, {Dai}, \&
  {Colpi}}]{Pfister+20}
{Pfister}, H., {Volonteri}, M., {Dai}, J.~L., \& {Colpi}, M. 2020, \mnras, 497,
  2276, \dodoi{10.1093/mnras/staa1962}

\bibitem[{{Phinney}(1989)}]{Phinney1989}
{Phinney}, E.~S. 1989, in IAU Symposium, Vol. 136, The Center of the Galaxy,
  ed. M.~{Morris}, 543, \dodoi{10.1017/S0074180900187054}

\bibitem[{Piran {et~al.}(2015)Piran, Svirski, Krolik, Cheng, \&
  Shiokawa}]{Piran+15}
Piran, T., Svirski, G., Krolik, J., Cheng, R.~M., \& Shiokawa, H. 2015, \apj,
  806, 164, \dodoi{10.1088/0004-637x/806/2/164}

\bibitem[{{Planck Collaboration} {et~al.}(2020){Planck Collaboration},
  {Aghanim}, {Akrami}, {Ashdown}, {Aumont}, {Baccigalupi}, {Ballardini},
  {Banday}, {Barreiro}, {Bartolo}, {Basak}, {Battye}, {Benabed}, {Bernard},
  {Bersanelli}, {Bielewicz}, {Bock}, {Bond}, {Borrill}, {Bouchet}, {Boulanger},
  {Bucher}, {Burigana}, {Butler}, {Calabrese}, {Cardoso}, {Carron},
  {Challinor}, {Chiang}, {Chluba}, {Colombo}, {Combet}, {Contreras}, {Crill},
  {Cuttaia}, {de Bernardis}, {de Zotti}, {Delabrouille}, {Delouis}, {Di
  Valentino}, {Diego}, {Dor{\'e}}, {Douspis}, {Ducout}, {Dupac}, {Dusini},
  {Efstathiou}, {Elsner}, {En{\ss}lin}, {Eriksen}, {Fantaye}, {Farhang},
  {Fergusson}, {Fernandez-Cobos}, {Finelli}, {Forastieri}, {Frailis},
  {Fraisse}, {Franceschi}, {Frolov}, {Galeotta}, {Galli}, {Ganga},
  {G{\'e}nova-Santos}, {Gerbino}, {Ghosh}, {Gonz{\'a}lez-Nuevo}, {G{\'o}rski},
  {Gratton}, {Gruppuso}, {Gudmundsson}, {Hamann}, {Handley}, {Hansen},
  {Herranz}, {Hildebrandt}, {Hivon}, {Huang}, {Jaffe}, {Jones}, {Karakci},
  {Keih{\"a}nen}, {Keskitalo}, {Kiiveri}, {Kim}, {Kisner}, {Knox},
  {Krachmalnicoff}, {Kunz}, {Kurki-Suonio}, {Lagache}, {Lamarre}, {Lasenby},
  {Lattanzi}, {Lawrence}, {Le Jeune}, {Lemos}, {Lesgourgues}, {Levrier},
  {Lewis}, {Liguori}, {Lilje}, {Lilley}, {Lindholm}, {L{\'o}pez-Caniego},
  {Lubin}, {Ma}, {Mac{\'\i}as-P{\'e}rez}, {Maggio}, {Maino}, {Mandolesi},
  {Mangilli}, {Marcos-Caballero}, {Maris}, {Martin}, {Martinelli},
  {Mart{\'\i}nez-Gonz{\'a}lez}, {Matarrese}, {Mauri}, {McEwen}, {Meinhold},
  {Melchiorri}, {Mennella}, {Migliaccio}, {Millea}, {Mitra},
  {Miville-Desch{\^e}nes}, {Molinari}, {Montier}, {Morgante}, {Moss}, {Natoli},
  {N{\o}rgaard-Nielsen}, {Pagano}, {Paoletti}, {Partridge}, {Patanchon},
  {Peiris}, {Perrotta}, {Pettorino}, {Piacentini}, {Polastri}, {Polenta},
  {Puget}, {Rachen}, {Reinecke}, {Remazeilles}, {Renzi}, {Rocha}, {Rosset},
  {Roudier}, {Rubi{\~n}o-Mart{\'\i}n}, {Ruiz-Granados}, {Salvati}, {Sandri},
  {Savelainen}, {Scott}, {Shellard}, {Sirignano}, {Sirri}, {Spencer},
  {Sunyaev}, {Suur-Uski}, {Tauber}, {Tavagnacco}, {Tenti}, {Toffolatti},
  {Tomasi}, {Trombetti}, {Valenziano}, {Valiviita}, {Van Tent}, {Vibert},
  {Vielva}, {Villa}, {Vittorio}, {Wandelt}, {Wehus}, {White}, {White},
  {Zacchei}, \& {Zonca}}]{Planck2018}
{Planck Collaboration}, {Aghanim}, N., {Akrami}, Y., {et~al.} 2020, \aap, 641,
  A6, \dodoi{10.1051/0004-6361/201833910}

\bibitem[{{Psaltis} \& {Johannsen}(2012)}]{JP11}
{Psaltis}, D., \& {Johannsen}, T. 2012, \apj, 745, 1,
  \dodoi{10.1088/0004-637X/745/1/1}

\bibitem[{{Rees}(1988)}]{Rees1988}
{Rees}, M.~J. 1988, \nat, 333, 523, \dodoi{10.1038/333523a0}

\bibitem[{{Remillard} \& {McClintock}(2006)}]{Remillard2006}
{Remillard}, R.~A., \& {McClintock}, J.~E. 2006, \araa, 44, 49,
  \dodoi{10.1146/annurev.astro.44.051905.092532}

\bibitem[{{Robertson} {et~al.}(2015){Robertson}, {Gallo}, {Zoghbi}, \&
  {Fabian}}]{Robertson15}
{Robertson}, D.~R.~S., {Gallo}, L.~C., {Zoghbi}, A., \& {Fabian}, A.~C. 2015,
  \mnras, 453, 3455, \dodoi{10.1093/mnras/stv1575}

\bibitem[{{Roming} {et~al.}(2005){Roming}, {Kennedy}, {Mason}, {Nousek}, {Ahr},
  {Bingham}, {Broos}, {Carter}, {Hancock}, {Huckle}, {Hunsberger}, {Kawakami},
  {Killough}, {Koch}, {McLelland}, {Smith}, {Smith}, {Soto}, {Boyd},
  {Breeveld}, {Holland}, {Ivanushkina}, {Pryzby}, {Still}, \&
  {Stock}}]{Roming05}
{Roming}, P. W.~A., {Kennedy}, T.~E., {Mason}, K.~O., {et~al.} 2005, \ssr, 120,
  95, \dodoi{10.1007/s11214-005-5095-4}

\bibitem[{{Roth}(1952)}]{Roth1952}
{Roth}, H. 1952, \aj, 57, 42, \dodoi{10.1086/106702}

\bibitem[{Roth {et~al.}(2016)Roth, Kasen, Guillochon, \&
  Ramirez-Ruiz}]{Roth+16}
Roth, N., Kasen, D., Guillochon, J., \& Ramirez-Ruiz, E. 2016, \apj, 827, 3,
  \dodoi{10.3847/0004-637x/827/1/3}

\bibitem[{{Roth} {et~al.}(2020){Roth}, {Rossi}, {Krolik}, {Piran}, {Mockler},
  \& {Kasen}}]{Roth+20}
{Roth}, N., {Rossi}, E.~M., {Krolik}, J., {et~al.} 2020, \ssr, 216, 114,
  \dodoi{10.1007/s11214-020-00735-1}

\bibitem[{{Ryu} {et~al.}(2020){Ryu}, {Krolik}, \& {Piran}}]{Ryu2020}
{Ryu}, T., {Krolik}, J., \& {Piran}, T. 2020, \apj, 904, 73,
  \dodoi{10.3847/1538-4357/abbf4d}

\bibitem[{{Ryu} {et~al.}(2023){Ryu}, {Krolik}, {Piran}, {Noble}, \&
  {Avara}}]{Ryu+23}
{Ryu}, T., {Krolik}, J., {Piran}, T., {Noble}, S.~C., \& {Avara}, M. 2023,
  \apj, 957, 12, \dodoi{10.3847/1538-4357/acf5de}

\bibitem[{{Sadowsk}(2009)}]{Sadowski2009A}
{Sadowsk}, A. 2009, \apjs, 183, 171, \dodoi{10.1088/0067-0049/183/2/171}

\bibitem[{{Sarin} \& {Metzger}(2023)}]{Sarin2023}
{Sarin}, N., \& {Metzger}, B.~D. 2023, arXiv e-prints, arXiv:2307.15121,
  \dodoi{10.48550/arXiv.2307.15121}

\bibitem[{{Saxton} {et~al.}(2017){Saxton}, {Read}, {Komossa}, {Lira},
  {Alexander}, \& {Wieringa}}]{Saxton2017}
{Saxton}, R.~D., {Read}, A.~M., {Komossa}, S., {et~al.} 2017, \aap, 598, A29,
  \dodoi{10.1051/0004-6361/201629015}

\bibitem[{{Schlegel} {et~al.}(1998){Schlegel}, {Finkbeiner}, \&
  {Davis}}]{Schlegel1998}
{Schlegel}, D.~J., {Finkbeiner}, D.~P., \& {Davis}, M. 1998, \apj, 500, 525,
  \dodoi{10.1086/305772}

\bibitem[{{Sesana} {et~al.}(2009){Sesana}, {Vecchio}, \&
  {Volonteri}}]{Sesana+09}
{Sesana}, A., {Vecchio}, A., \& {Volonteri}, M. 2009, \mnras, 394, 2255,
  \dodoi{10.1111/j.1365-2966.2009.14499.x}

\bibitem[{{Shakura} \& {Sunyaev}(1973)}]{Shakura1973}
{Shakura}, N.~I., \& {Sunyaev}, R.~A. 1973, \aap, 24, 337

\bibitem[{{Shimura} \& {Takahara}(1995)}]{Shimura1995}
{Shimura}, T., \& {Takahara}, F. 1995, \apj, 445, 780, \dodoi{10.1086/175740}

\bibitem[{Shiokawa {et~al.}(2015)Shiokawa, Krolik, Cheng, Piran, \&
  Noble}]{Shiokawa+15}
Shiokawa, H., Krolik, J.~H., Cheng, R.~M., Piran, T., \& Noble, S.~C. 2015,
  \apj, 804, 85, \dodoi{10.1088/0004-637x/804/2/85}

\bibitem[{{Shu} {et~al.}(2020){Shu}, {Zhang}, {Li}, {Jiang}, {Dou}, {Yan},
  {Xie}, {Shen}, {Sun}, {Liu}, \& {Wang}}]{Shu2020}
{Shu}, X., {Zhang}, W., {Li}, S., {et~al.} 2020, Nature Communications, 11,
  5876, \dodoi{10.1038/s41467-020-19675-z}

\bibitem[{{S{\k{a}}dowski}(2009)}]{Sadowski09}
{S{\k{a}}dowski}, A. 2009, \apjs, 183, 171, \dodoi{10.1088/0067-0049/183/2/171}

\bibitem[{{Steinberg} \& {Stone}(2022)}]{Steinberg+22}
{Steinberg}, E., \& {Stone}, N.~C. 2022, arXiv e-prints, arXiv:2206.10641,
  \dodoi{10.48550/arXiv.2206.10641}

\bibitem[{{Stone} \& {Metzger}(2016)}]{Stone2016}
{Stone}, N.~C., \& {Metzger}, B.~D. 2016, \mnras, 455, 859,
  \dodoi{10.1093/mnras/stv2281}

\bibitem[{{Stone} {et~al.}(2020){Stone}, {Vasiliev}, {Kesden}, {Rossi},
  {Perets}, \& {Amaro-Seoane}}]{Stone+20}
{Stone}, N.~C., {Vasiliev}, E., {Kesden}, M., {et~al.} 2020, \ssr, 216, 35,
  \dodoi{10.1007/s11214-020-00651-4}

\bibitem[{{Str{\"u}der} {et~al.}(2001){Str{\"u}der}, {Briel}, {Dennerl},
  {Hartmann}, {Kendziorra}, {Meidinger}, {Pfeffermann}, {Reppin}, {Aschenbach},
  {Bornemann}, {Br{\"a}uninger}, {Burkert}, {Elender}, {Freyberg}, {Haberl},
  {Hartner}, {Heuschmann}, {Hippmann}, {Kastelic}, {Kemmer}, {Kettenring},
  {Kink}, {Krause}, {M{\"u}ller}, {Oppitz}, {Pietsch}, {Popp}, {Predehl},
  {Read}, {Stephan}, {St{\"o}tter}, {Tr{\"u}mper}, {Holl}, {Kemmer}, {Soltau},
  {St{\"o}tter}, {Weber}, {Weichert}, {von Zanthier}, {Carathanassis}, {Lutz},
  {Richter}, {Solc}, {B{\"o}ttcher}, {Kuster}, {Staubert}, {Abbey}, {Holland},
  {Turner}, {Balasini}, {Bignami}, {La Palombara}, {Villa}, {Buttler},
  {Gianini}, {Lain{\'e}}, {Lumb}, \& {Dhez}}]{Struder+01}
{Str{\"u}der}, L., {Briel}, U., {Dennerl}, K., {et~al.} 2001, \aap, 365, L18,
  \dodoi{10.1051/0004-6361:20000066}

\bibitem[{{Sun} {et~al.}(2018){Sun}, {Grier}, \& {Peterson}}]{Sun2018}
{Sun}, M., {Grier}, C.~J., \& {Peterson}, B.~M. 2018, {PyCCF: Python Cross
  Correlation Function for reverberation mapping studies}, Astrophysics Source
  Code Library, record ascl:1805.032.
\newblock \doeprint{1805.032}

\bibitem[{{Teboul} {et~al.}(2024){Teboul}, {Stone}, \& {Ostriker}}]{Teboul2024}
{Teboul}, O., {Stone}, N.~C., \& {Ostriker}, J.~P. 2024, \mnras, 527, 3094,
  \dodoi{10.1093/mnras/stad3301}

\bibitem[{{Timmer} \& {Koenig}(1995)}]{Timmer1995}
{Timmer}, J., \& {Koenig}, M. 1995, \aap, 300, 707

\bibitem[{{Turner} {et~al.}(2001){Turner}, {Abbey}, {Arnaud}, {Balasini},
  {Barbera}, {Belsole}, {Bennie}, {Bernard}, {Bignami}, {Boer}, {Briel},
  {Butler}, {Cara}, {Chabaud}, {Cole}, {Collura}, {Conte}, {Cros}, {Denby},
  {Dhez}, {Di Coco}, {Dowson}, {Ferrando}, {Ghizzardi}, {Gianotti}, {Goodall},
  {Gretton}, {Griffiths}, {Hainaut}, {Hochedez}, {Holland}, {Jourdain},
  {Kendziorra}, {Lagostina}, {Laine}, {La Palombara}, {Lortholary}, {Lumb},
  {Marty}, {Molendi}, {Pigot}, {Poindron}, {Pounds}, {Reeves}, {Reppin},
  {Rothenflug}, {Salvetat}, {Sauvageot}, {Schmitt}, {Sembay}, {Short},
  {Spragg}, {Stephen}, {Str{\"u}der}, {Tiengo}, {Trifoglio}, {Tr{\"u}mper},
  {Vercellone}, {Vigroux}, {Villa}, {Ward}, {Whitehead}, \&
  {Zonca}}]{Turner+01}
{Turner}, M.~J.~L., {Abbey}, A., {Arnaud}, M., {et~al.} 2001, \aap, 365, L27,
  \dodoi{10.1051/0004-6361:20000087}

\bibitem[{{Ulmer}(1999)}]{Ulmer99}
{Ulmer}, A. 1999, \apj, 514, 180, \dodoi{10.1086/306909}

\bibitem[{{Uttley} {et~al.}(2014){Uttley}, {Cackett}, {Fabian}, {Kara}, \&
  {Wilkins}}]{2014A&ARv..22...72U}
{Uttley}, P., {Cackett}, E.~M., {Fabian}, A.~C., {Kara}, E., \& {Wilkins},
  D.~R. 2014, \aapr, 22, 72, \dodoi{10.1007/s00159-014-0072-0}

\bibitem[{{van den Bosch} {et~al.}(2012){van den Bosch}, {Gebhardt},
  {G{\"u}ltekin}, {van de Ven}, {van der Wel}, \& {Walsh}}]{vandenBosch+12}
{van den Bosch}, R. C.~E., {Gebhardt}, K., {G{\"u}ltekin}, K., {et~al.} 2012,
  \nat, 491, 729, \dodoi{10.1038/nature11592}

\bibitem[{van Velzen {et~al.}(2011)van Velzen, Körding, \&
  Falcke}]{vanVelzen+11}
van Velzen, S., Körding, E., \& Falcke, H. 2011, Monthly Notices of the Royal
  Astronomical Society: Letters, 417, L51,
  \dodoi{10.1111/j.1745-3933.2011.01118.x}

\bibitem[{van Velzen {et~al.}(2019)van Velzen, Stone, Metzger, Gezari, Brown,
  \& Fruchter}]{vanVelzen+19}
van Velzen, S., Stone, N.~C., Metzger, B.~D., {et~al.} 2019, \apj, 878, 82,
  \dodoi{10.3847/1538-4357/ab1844}

\bibitem[{{van Velzen} {et~al.}(2021){van Velzen}, {Gezari}, {Hammerstein},
  {Roth}, {Frederick}, {Ward}, {Hung}, {Cenko}, {Stein}, {Perley}, {Taggart},
  {Foley}, {Sollerman}, {Blagorodnova}, {Andreoni}, {Bellm}, {Brinnel}, {De},
  {Dekany}, {Feeney}, {Fremling}, {Giomi}, {Golkhou}, {Graham}, {Ho},
  {Kasliwal}, {Kilpatrick}, {Kulkarni}, {Kupfer}, {Laher}, {Mahabal}, {Masci},
  {Miller}, {Nordin}, {Riddle}, {Rusholme}, {van Santen}, {Sharma}, {Shupe}, \&
  {Soumagnac}}]{van2021ApJ}
{van Velzen}, S., {Gezari}, S., {Hammerstein}, E., {et~al.} 2021, \apj, 908, 4,
  \dodoi{10.3847/1538-4357/abc258}

\bibitem[{{Vasiliev} {et~al.}(2015){Vasiliev}, {Antonini}, \&
  {Merritt}}]{Vasiliev2015}
{Vasiliev}, E., {Antonini}, F., \& {Merritt}, D. 2015, \apj, 810, 49,
  \dodoi{10.1088/0004-637X/810/1/49}

\bibitem[{{Wang} \& {Merritt}(2004)}]{Wang&Merritt2004}
{Wang}, J., \& {Merritt}, D. 2004, \apj, 600, 149, \dodoi{10.1086/379767}

\bibitem[{{Wegg} \& {Nate Bode}(2011)}]{Wegg2011}
{Wegg}, C., \& {Nate Bode}, J. 2011, \apjl, 738, L8,
  \dodoi{10.1088/2041-8205/738/1/L8}

\bibitem[{{Welsh}(1999)}]{Welsh1999}
{Welsh}, W.~F. 1999, \pasp, 111, 1347, \dodoi{10.1086/316457}

\bibitem[{{Wen} {et~al.}(2023){Wen}, {Jonker}, {Stone}, {Van Velzen}, \&
  {Zabludoff}}]{Wen22b}
{Wen}, S., {Jonker}, P.~G., {Stone}, N.~C., {Van Velzen}, S., \& {Zabludoff},
  A.~I. 2023, \mnras, 522, 1155, \dodoi{10.1093/mnras/stad991}

\bibitem[{Wen {et~al.}(2021)Wen, Jonker, Stone, \& Zabludoff}]{Wen_2021}
Wen, S., Jonker, P.~G., Stone, N.~C., \& Zabludoff, A.~I. 2021, \apj, 918, 46,
  \dodoi{10.3847/1538-4357/ac00b5}

\bibitem[{{Wen} {et~al.}(2022){Wen}, {Jonker}, {Stone}, {Zabludoff}, \&
  {Cao}}]{Wen_2022}
{Wen}, S., {Jonker}, P.~G., {Stone}, N.~C., {Zabludoff}, A.~I., \& {Cao}, Z.
  2022, \apj, 933, 31, \dodoi{10.3847/1538-4357/ac70c5}

\bibitem[{{Wen} {et~al.}(2020){Wen}, {Jonker}, {Stone}, {Zabludoff}, \&
  {Psaltis}}]{Wen2020}
{Wen}, S., {Jonker}, P.~G., {Stone}, N.~C., {Zabludoff}, A.~I., \& {Psaltis},
  D. 2020, \apj, 897, 80, \dodoi{10.3847/1538-4357/ab9817}

\bibitem[{{Wevers}(2020)}]{Wevers20}
{Wevers}, T. 2020, \mnras, 497, L1, \dodoi{10.1093/mnrasl/slaa097}

\bibitem[{{Wevers} {et~al.}(2019){Wevers}, {Pasham}, {van Velzen}, {Leloudas},
  {Schulze}, {Miller-Jones}, {Jonker}, {Gromadzki}, {Kankare}, {Hodgkin},
  {Wyrzykowski}, {Kostrzewa-Rutkowska}, {Moran}, {Berton}, {Maguire}, {Onori},
  {Mattila}, \& {Nicholl}}]{Wevers_2019}
{Wevers}, T., {Pasham}, D.~R., {van Velzen}, S., {et~al.} 2019, \mnras, 488,
  4816, \dodoi{10.1093/mnras/stz1976}

\bibitem[{Wevers {et~al.}(2021)Wevers, Pasham, van Velzen, Miller-Jones,
  Uttley, Gendreau, Remillard, Arzoumanian, Löwenstein, \&
  Chiti}]{Wevers_2021}
Wevers, T., Pasham, D.~R., van Velzen, S., {et~al.} 2021, \apj, 912, 151,
  \dodoi{10.3847/1538-4357/abf5e2}

\bibitem[{Wevers {et~al.}(2023)Wevers, Coughlin, Pasham, Guolo, Sun, Wen,
  Jonker, Zabludoff, Malyali, Arcodia, Liu, Merloni, Rau, Grotova, Short, \&
  Cao}]{Wevers_2023}
Wevers, T., Coughlin, E.~R., Pasham, D.~R., {et~al.} 2023, The Astrophysical
  Journal Letters, 942, L33, \dodoi{10.3847/2041-8213/ac9f36}

\bibitem[{{Yao} {et~al.}(2022){Yao}, {Lu}, {Guolo}, {Pasham}, {Gezari},
  {Gilfanov}, {Gendreau}, {Harrison}, {Cenko}, {Kulkarni}, {Miller}, {Walton},
  {Garc{\'\i}a}, {van Velzen}, {Alexander}, {Miller-Jones}, {Nicholl},
  {Hammerstein}, {Medvedev}, {Stern}, {Ravi}, {Sunyaev}, {Bloom}, {Graham},
  {Kool}, {Mahabal}, {Masci}, {Purdum}, {Rusholme}, {Sharma}, {Smith}, \&
  {Sollerman}}]{Yao2022ApJ}
{Yao}, Y., {Lu}, W., {Guolo}, M., {et~al.} 2022, \apj, 937, 8,
  \dodoi{10.3847/1538-4357/ac898a}

\bibitem[{{Yao} {et~al.}(2023){Yao}, {Ravi}, {Gezari}, {van Velzen}, {Lu},
  {Schulze}, {Somalwar}, {Kulkarni}, {Hammerstein}, {Nicholl}, {Graham},
  {Perley}, {Cenko}, {Stein}, {Ricarte}, {Chadayammuri}, {Quataert}, {Bellm},
  {Bloom}, {Dekany}, {Drake}, {Groom}, {Mahabal}, {Prince}, {Riddle},
  {Rusholme}, {Sharma}, {Sollerman}, \& {Yan}}]{Yao+23}
{Yao}, Y., {Ravi}, V., {Gezari}, S., {et~al.} 2023, \apjl, 955, L6,
  \dodoi{10.3847/2041-8213/acf216}

\bibitem[{{Zanazzi} \& {Ogilvie}(2020)}]{Zanazzi+20}
{Zanazzi}, J.~J., \& {Ogilvie}, G.~I. 2020, \mnras, 499, 5562,
  \dodoi{10.1093/mnras/staa3127}

\bibitem[{{Zdziarski} {et~al.}(2020){Zdziarski}, {Szanecki}, {Poutanen},
  {Gierli{\'n}ski}, \& {Biernacki}}]{Zdziarski_2020}
{Zdziarski}, A.~A., {Szanecki}, M., {Poutanen}, J., {Gierli{\'n}ski}, M., \&
  {Biernacki}, P. 2020, \mnras, 492, 5234, \dodoi{10.1093/mnras/staa159}

\bibitem[{{Zhang}(2022)}]{Zhang2022}
{Zhang}, W. 2022, \mnras, 511, 19, \dodoi{10.1093/mnrasl/slab133}

\end{thebibliography}
\bibliographystyle{aasjournal}

\end{document}